\documentclass[a4paper,12pt,reqno]{amsart}
\usepackage{latexsym}
\usepackage{amssymb,color}

\newtheorem{thm}{Theorem}[section]
\newtheorem{cor}[thm]{Corollary}

\newtheorem{prop}[thm]{Proposition}

\theoremstyle{remark}
\newtheorem{rem}[thm]{Remark}
\numberwithin{equation}{section} \makeatletter
\def\@cite#1#2{#1\if@tempswa , #2\fi}
\def\@biblabel#1{$^{\hbox{\scriptsize{#1}}}$}
\makeatother
\newcommand{\R}{\mathbb R}

\newcommand{\To}{\longrightarrow}

\newcommand{\h}{\mathcal{H}}
\newcommand{\N}{\mathbb N}

\newcommand{\pr}{^{\prime}}

\newcommand{\beq}{\begin{equation}}
\newcommand{\eeq}{\end{equation}}
\newcommand{\ben}{\begin{enumerate}}
\newcommand{\een}{\end{enumerate}}

\newcommand{\C}{\mathbb C}

\newcommand{\p}{\textit{\textbf{p}}}

\font\sbi=cmmib10 \font\bi=cmmib10 scaled \magstep1
\usepackage{mathtools}
\begin{document}

\title[]{Solution of a linearized model of\\
Heisenberg's fundamental equation II}

\author{E. Br\"uning \and S. Nagamachi }

\address[E. Br\"uning]{School of Mathematical Sciences, University of KwaZulu-Natal, Private Bag X54001,
Durban 4000, South Africa} \email{bruninge@ukzn.ac.za}

\address[S. Nagamachi]{
Department of Applied Physics and Mathematics, Faculty of
Engineering, The University of Tokushima\\ Tokushima 770-8506,
Japan} \email{shigeaki@pm.tokushima-u.ac.jp}

\subjclass{81T05, 32A45, 46F15} \keywords{quantum field theory,
tempered ultra-hyperfunctions, quantum fields with fundamental
length}
\begin{abstract}
We propose to look at (a simplified version of) Heisenberg's fundamental field equation
(see [\cite{He66}]) as a relativistic quantum field theory with a fundamental length,
as introduced in [\cite{BN04}] and give a solution in terms of Wick power series of free
fields which converge in the sense of ultrahyperfunctions but not in the sense of distributions.

The solution of this model has been prepared in [\cite{NB07a}] by calculating all $n$-point functions using path integral quantization. The functional representation derived in this part is essential for the verification of our condition of extended causality.
The verification of the remaining defining conditions of a relativistic quantum field theory is much simpler
through the use of Wick power series. Accordingly in this second part we use Wick power series techniques to define our basic fields and derive their properties.
\end{abstract}

\maketitle \tableofcontents

\section{Introduction}
\subsection{Motivation and outline of paper}
\noindent
 Heisenberg's fundamental field equation (see [\cite{He66}])
\beq \label{eq:ffeq}
  \gamma _{\mu } \frac{\partial }{\partial x_{\mu }}\psi (x) \pm  l^{2} \gamma _{\mu }\gamma _{5} :\psi (x) \bar{\psi } (x) \gamma ^{\mu } \gamma _{5}\psi (x): \  = 0
\eeq
contains a parameter $l$ of the dimension of length and accordingly
one might speculate that this parameter can play the r\^ole of the
fundamental length of a quantum field theory with a fundamental length as
introduced in [\cite{BN04}].
Unfortunately, nobody knows to solve this equation. However there is
a simplification of Heisenberg's equation which is solvable in the
sense of classical field theory, namely the system of equations
\begin{equation} \label{eq:soluble}
\left\{
\begin{array}{l}
 (\Box  + m^{2})\phi (x) = 0 \\
   \displaystyle \left(  i\gamma _{\mu }\frac{\partial }{\partial x_{\mu }} - M \right)
   \psi (x) = -2l^{2} \gamma _{\mu } :\psi (x)\phi (x) \frac{\partial \phi (x)}{\partial x_{\mu }}:
\end{array}
\right.
\end{equation}
for a Klein-Gordon field $\phi $ and a spinor field $\psi $. It is
this system of coupled equations which we discuss in the framework
of [\cite{BN04}]. In [\cite{NB07a}], this system with the Lagrangian density
$$       L(x) = L_{Ff}(x) + L_{Fb}(x) + L_{I}(x), \  L_{Ff}(x) = \bar{\psi }  (x) (i\gamma ^{\mu }\partial _{\mu } -\tilde{m}  )\psi (x),$$
$$       L_{Fb}(x) = \frac{1}{2}\{ (\partial _{\mu }\phi (x))^{2} - m^{2}\phi (x)^{2}\} , \  L_{I}(x) = 2l^{2}(\bar{\psi }  (x)\gamma ^{\mu }\psi (x))\phi (x)\partial _{\mu }\phi (x)$$
is quantized by the method of path integral, that is, the
$n$-point Schwinger functions are calculated by the Euclideanized
lattice approximation (with infinitesimal spacing, in the framework of nonstandard analysis) of following
path integral:
$$       \int \prod _{j=1}^{n} \psi ^{r_{j}}_{\alpha _{j}}(x_{j}) \exp i \left\{  \int _{ \mathbb R^{4}} L_{I}(x) dx\right\}  d{\mathcal D}  (\psi , \bar{\psi }  ) d{\mathcal G}  (\phi )$$
$$       \times  \left\{ \int  \exp i \left\{  \int _{ \mathbb R^{4}} L_{I}(x) dx\right\}  d{\mathcal D}  (\psi , \bar{\psi }  ) d{\mathcal G}  (\phi )\right\} ^{-1},$$
$$       d{\mathcal G}  (\phi ) = \exp i \left\{  \int _{ \mathbb R^{4}} L_{Fb}(x) dx \right\}  \prod _{x\in  \mathbb R^{4}} d\phi (x)$$
$$       d{\mathcal D}  (\psi , \bar{\psi }  ) = \exp i\left\{ \int _{ \mathbb R^{4}} L_{Ff}(x) dx\right\}  \prod _{x\in  \mathbb R^{4}}\prod _{\alpha =1}^{4}\psi _{\alpha }(x)\bar{\psi }  _{\alpha }(x),$$
where $\psi ^{1} = \psi $, $\psi ^{2} = \bar{\psi }  $.

After  renormalization we obtain the continuous limit of the
lattice Schwinger functions.  Then the Wightman functions are obtained by
Wick rotation of the Schwinger functions.  If these Wightman
functions satisfy the axioms of the relativistic quantum field
theory, then, by the reconstruction theorem, we can construct the
operator valued generalized functions $\phi (x)$ and $\psi (x)$.  We understand
that these fields $\phi (x)$ and $\psi (x)$ are the solutions of the system
defined by the Lagrangian density (1.3) according to standard interpretation
of renormalization procedure.

  In this paper, we try to construct the quantum fields $\phi (x)$ and $\psi (x)$ which satisfy the system of differential equations (1.2), then show that these
fields satisfy the axioms of the relativistic quantum field theory.
In [5], it is shown that the Wightman functions of $\psi(x)$ are not
tempered distributions used in the usual Wightman axioms but
tempered ultra-hyperfunctions which are used to formulate the
quantum field theory with a fundamental length in [1].

In Section 2 we show that the $n$-point functionals constructed in this way satisfy the spinor version of the functional characterization of our condition of extended causality of [\cite{BN04}]. In order to verify the remaining
defining conditions of our relativistic field theory  with a fundamental length we use Wick power series
to define the theory. Accordingly, in this second part
 we construct an operator valued generalized
function $\psi(x)$ satisfying (\ref{eq:soluble}).  The basic idea to
solve the system (\ref{eq:soluble}) is quite natural: \\
Take a Klein-Gordon field of mass $m$ and suppose that we can show the following
three statements:
\begin{enumerate}
\item[A)] the Wick power series
\beq \label{eq:Wps}
       \rho(x)= :e^{il^{2}\phi(x)^{2}}: = \sum_{n=0}^{\infty} i^{n} l^{2n}
        :\phi(x)^{2n}: / n!  \eeq
and
$$   \rho ^{*}(x) = :e^{-il^{2}\phi (x)^{2}}: = \sum _{n=0}^{\infty } (-i)^{n} l^{2n}:\phi (x)^{2n}: / n!$$
are well-defined as an operator-valued ultra-hyperfunctions.
\item[B)] $\rho (x)$ satisfies
\begin{multline} \label{eq:Wpsder}
 \frac{\partial }{\partial x^{\mu }} \rho (x) = 2il^{2} :e^{il^{2}\phi (x)^{2}}
  \phi (x) \frac{\partial }{\partial x^{\mu }} \phi (x): \\= 2il^{2} :\rho (x) \phi (x)
  \frac{\partial }{\partial x^{\mu }} \phi(x):.\end{multline}
\item[C)] the free Dirac field $\psi _{0}(x)$ is a multiplier for the field
$\rho $ and so, define the
field \beq \label{eq:field} \psi (x) = \psi _{0}(x) \rho (x), \eeq
and calculate
\begin{multline*}
   \left( i \gamma_{\mu }\frac{\partial }{\partial x^{\mu }} - M\right) \psi (x)\\
 = \left[ \left( i \gamma _{\mu }\frac{\partial }{\partial x^{\mu }} - M\right) \psi_{0}(x) \right]
 \rho (x)  + \gamma _{\mu }\psi _{0} (x) \frac{\partial }{\partial x^{\mu }}\rho (x)\\
        = -2l^{2} \gamma_{\mu }\psi_{0}(x) :\rho (x) \phi(x)
        \frac{\partial}{\partial x^{\mu }} \phi(x): \\
        = -2l^{2} \gamma_{\mu } : \psi(x) {\phi}(x) \frac{\partial}{\partial x^{\mu}} \phi(x):.
\end{multline*}

\end{enumerate}
Thus, if A) -- C) hold, the operator-valued ultra-hyperfunction
$\psi (x)$ satisfies Equation (\ref{eq:soluble}).\newline
\indent In [\cite{BN04}] statement A) is shown together with the fact that
the fields $\phi (x)$, $\rho (x)$ and $\rho^{*}(x)$ satisfy the
axioms of ultra-hyperfunction quantum field theory (UHFQFT).
In Section 3 the convergence of the Wick power series for $\rho(x)=:e^{g\phi(x)^2}:$ is recalled form [\cite{BN04}]. In the next section the important differential equation $\partial_{\mu}\rho(x)=2i\ell^2 :\rho(x)\phi(x)\partial_{\mu}\phi(x):$ is proven.
Then in order to prepare the treatment of Dirac fields, in Section 5 the axioms of UHFQFT with a fundamental length $\ell $, for general type of (in particular spinor) fields are presented.  In order to show statement C), we study some properties of $\rho (x)$ which follow from the axioms of UHFQFT in Section 5.  In Section 7 it is shown that the pointwise product (\ref{eq:field}) of two operator-valued tempered
ultrahyperfunctions is well-defined and thus statement  C)
can be established; and it is shown that $\phi (x)$, $\psi (x) = \psi
_{0}(x)\rho (x)$ and $\bar{\psi } (x) = \rho ^{*}(x)\bar{\psi }
_{0}(x) = \bar{\psi } _{0}(x)\rho ^{*}(x)$ satisfy all axioms of
UHFQFT for general type fields as presented in Section 4, and their Wightman functions are the same ones obtained in [5] using path integral methods.

\subsection{Localization properties of tempered ultra-hyperfunctions}
As announced, in Section 2 we are going to show that the system of $n$-point functionals as constructed in the first part satisfy the condition of extended causality. Since this condition is based on the localization properties of tempered ultra-hyperfunctions we explain here briefly the technical realization of these
localization properties. To simplify matters we use a simple one-dimensional model first.

Denote $T(-\ell , \ell ) = \R + i(-\ell , \ell ), T[-k, k] = \R + i[-k, k] \subset \C
$, and let ${\mathcal T} (T(-\ell , \ell ))$ be the set of functions $f$ holomorphic in
$T(-\ell , \ell ) $ and rapidly decreasing in any $T[-k, k] \subset  T(-\ell , \ell )$.
Then for $\vert a\vert < \ell $, we get
$$ \int_{-\infty}^{\infty} \sum_{n = 0}^{\infty} \frac{a^{n}}{n!} \delta  ^{(n)}(x) f(x) dx = \sum_{n = 0}^{\infty } \frac{(-a)^{n}}{n!} f^{(n)}(0)$$
$$ = f(- a) = \int_{-\infty}^{\infty} \delta (x + a) f(x) dx.$$
The above equality implies the following two facts.
\begin{enumerate}
\item[(A)]
If $\vert a\vert < \ell $ then $\Delta_{N}(x) = \sum_{n = 0}^{N} \frac{a^{n}}{n!} \delta ^{(n)}(x)$ converges to $\delta (x + a) =
\delta _{-a}(x)$ in ${\mathcal T} (T(-\ell , \ell ))^{\prime }$ as $N \rightarrow  \infty $. Clearly, for all $N \in  \N$,
${\rm supp\, }\Delta_{N} = \{ 0\} $ while for the limit we find ${\rm supp\, }\delta_{-a} = \{ -a\} $.
\item[(B)] If $\vert a\vert > \ell $, $\Delta_{N}(x)$ does not converge in ${\mathcal T} (T(-\ell, \ell
)^{\prime }$.
\end{enumerate}
(A) and (B) say: Elements in ${\mathcal T} (T(-\ell , \ell))^{\prime }$ do not allow to
distinguish between $\{ 0\} $ and $\{ -a\} $, if $\vert a\vert < \ell $, but if $\vert a\vert  > \ell $
then elements in ${\mathcal T} (T(-\ell , \ell ))^{\prime }$ can be used to distinguish between
the locations $\{ 0\} $ and $\{ -a\} $.  Such a length $\ell $ is considered to
be the fundamental length.  ${\mathcal T} (T(-\infty , \infty))^{\prime }$ is called the space of
the tempered ultrahyperfunctions, where ${\mathcal T}(T(-\infty , \infty))= \lim_{\infty  \leftarrow  \ell }
{\mathcal T} (T(-\ell , \ell ))$ is the space of rapidly decreasing entire functions.
${\mathcal T} (T(-\ell, \ell))^{\prime }$ is the space of tempered ultra-hyperfunctions whose
carrier are contained in $T(-\ell , \ell )$.  The standard locality condition of quantum field theory in terms of Schwartz distributions is extended
using the notion of carrier of analytic functionals
(functionals over the test-function space of analytic functions)
instead of the notion of support of Schwartz distributions.

For a field $\phi (x)$ satisfying the standard Wightman axioms, 
the two-point functional
$(\Phi , \phi (x)\phi (y)\Psi )$ is a functional over the
test-function space ${\mathcal S} (\R^{2\cdot 4})$, i.e., a
tempered distribution.  However, for the field $\psi (x)$
satisfying Equation (1.2), $(\Phi , \psi (x)\psi (y)\Psi )$ is not
a functional over the test-function space ${\mathcal S}
(\R^{2\cdot 4})$ but, as shown in sections 2 and 7 of this paper,
a functional over the test-function space ${\mathcal T} (T(L^{\ell
^{\prime}}))$ for any $\ell ^{\prime } > \ell  = \ell _{m}(l) = l/(\sqrt{2}\pi ) + O(l^{2})$,
where
$$T(L^{\ell^{\prime }}) = \R^{2\cdot 4} + i L^{\ell ^{\prime }}, \  L^{\ell ^{\prime }} = \{ (y_{1}, y_{2}) \in  \R^{2\cdot 4}; \vert y_{1} - y_{2}\vert  < \ell ^{\prime }\} .$$
Thus such a functional can distinguish two events occurring at $x_{1}$ and
$x_{2}$ if the distance between $x_{1}$ and $x_{2}$ is greater than $\ell $, and
cannot distinguish them if the distance is smaller than $\ell $.  In
that sense, the field $\psi (x)$ does not define a local field but
 a quasi-local field with a fundamental length $\ell $.

 It is quite interesting that the parameter $l$ with the dimension of length
contained in Equations (1.2) is essentially the fundamental length
$\ell ^{\prime } > \ell  = \ell _{m}(l) = l/(\sqrt{2}\pi ) + O(l^{2})$ in the sense of this theory.

\section{Verification of extended causality}
In this section we are going to prove that the system of functionals
 (5.7) of Part I (see [\cite{NB07a}]), i.e., the functionals on ${\mathcal T}(T(\R^{4n}))$
  \begin{equation}\label{eq:n-point}
 {\mathcal
W}^{r}_{\alpha}(f)=\int_{\prod_{j=1}^{n}\Gamma_{j}}(\det
A(z))^{-1/2} {\mathcal W}^{r}_{0,\alpha}(z_{1},\ldots,z_{n}) f(z)
dz,
\end{equation}
where $A(z)$ is the $n \times  n$ symmetric matrix whose entries
$a_{j,k}$ are given by
$$       a_{j, k} = a_{k, j} = 2h_{r_{j}}h_{r_{k}}l^{2}D^{(-)}_{m}(z_{j} - z_{k})$$
for $r_{j} = \pm 1$, $h_{\pm 1} = e^{\pm i\pi /4}$, $j < k$ and
$a_{j, j} = 1$, and where the paths $\Gamma_{j}$ are $\R^{4} + i(y_{j}^{0},0,0,0)$ for appropriately chosen constants $y_{j}^{0}$,  satisfies the spinor version of condition (R3) of extended locality as presented in [\cite{BN04}].
  For convenience we recall this condition here:\\[1mm]
{\bf(R3)} (Condition of extended causality): For all
$n=2,3,\ldots$ and all $j=1,\ldots,n-1$ denote
\begin{align*}
 L^{\ell }_j& = \{ x = (x_{1}, \ldots, x_{n}) \in  \R^{4n};
 \vert x_{j} - x_{j+1}\vert _1 < \ell \},\\
W^{\ell }_j &= \{ (z_{1}, \ldots , z_{n}) \in  \C^{4n}; z_{j} -
z_{j+1}
 \in  V^{\ell }\} ,
  \end{align*}
where \beq \label{cnbhd} V^{\ell }=\{z \in \C^{4}; \exists x \in
V;  \vert {\rm Re \,}z - x\vert <\ell, \  \vert{\rm Im \,}z
\vert_1 <\ell \}.\eeq is a complex neighborhood of light cone $V$.
Then, for any $\ell\pr > \ell $,
\begin{enumerate}
 \item[(i)] the
functional on ${\mathcal T}  (T(\R^{4n}))$
$${\mathcal T}  (T(\R^{4n})) \ni  f \rightarrow
{\mathcal W} ^{r}_{\alpha }(f) \in \C$$ is extended continuously
to ${\mathcal T} (T(L_{j}^{\ell ^{\prime }}))$, and \item[(ii)]
the functional on ${\mathcal T} (T(\R^{4n}))$
$$  f \rightarrow  {\mathcal W}^{r_{1} \ldots r_{j} r_{j+1} \ldots r_{n}}_{\alpha _{1} \ldots \alpha _{j} \alpha _{j+1} \ldots \alpha _{n}}(f)  +
{\mathcal W}^{r_{1} \ldots r_{j+1} r_{j+} \ldots r_{n}}_{\alpha
_{1} \ldots \alpha _{j+1} \alpha _{j} \ldots \alpha _{n}}(f) \in
\C$$ is extended continuously to ${\mathcal
T}(W_{j}^{\ell^{\prime}})$.
\end{enumerate}
\begin{rem}\label{comp-to-old}
 In our previous paper [\cite{BN04}], we defined a complex
neighbourhood $V^{\ell }$ by \beq \label{nbhd-old} V^{\ell }=\{z
\in \C^{4}; \exists x \in V; \vert {\rm Re \,}z - x\vert +
\vert{\rm Im \,}z \vert_1 <\ell \}.\eeq

But we found that to treat the present model, the neighbourhood
(\ref{cnbhd}) is convenient, and by this change of the
$\ell$-neighbourhood of $V$, our theory [\cite{BN04}] is
not affected.
\end{rem}
 In order to verify this condition  fix $j\in \{1,\ldots,n-1\}$ and assume
\begin{equation}\label{eq:dist-ell}
 y_{j+1}^{0}-y_{j}^{0} > \ell=l/(\sqrt{2}\pi),\end{equation}
 then by estimate (5.6) of Part I, i.e., the global estimate
\begin{equation}\label{estim-2point}
 \vert D_{m}^{(-)}(x^{0}-i\epsilon,\mbox{\bi x})\vert \leq (2\pi
\epsilon)^{-2} \quad \textrm{for all}\;x \in \R^{4}, \quad \forall\; \epsilon >0,
\end{equation}
 it follows
$$\vert 4l^{4}D_{m}^{(-)}(z_{j}-z_{j+1})^{2}\vert < 1.$$
Introduce
\begin{equation}\label{det-details}
 Q_{n,j}(a_{i,k})=\sum_{\stackrel{(i,k,\ldots,l)\neq(1,2,\ldots,n)}
{(i,k,\ldots,l)\neq(1,2,\ldots,j+1,j,\ldots,n)}}{\rm sgn\,}(i, k,
\ldots,l)a_{1,j}a_{2,k}\cdots a_{n,l}\end{equation} and denote by
$\sigma(j+1,j)$ the permutation $(1,\ldots,j-1,j,j+1,\ldots,n)\To
(1,\ldots,j-1,j+1,j,\ldots,n)$.
 Then we have
\begin{multline*}
 P_n(a_{i,k})={\rm sgn\,}(\sigma(j+1,j)))a_{1,1}a_{2,2}
 \cdots a_{j-1,j-1}a_{j,j+1}a_{j+1,j} a_{j+1,j+1} \cdots a_{n,n}\\+
  Q_{n,j}(a_{i,k})
=- a_{j,j+1}^{2}+Q_{n,j}(a_{i,k})=\pm 4l^{2}D_{m}^{(-)}(z_{j}-
z_{j+1})^{2} + Q_{n,j}(a_{i,k}).
\end{multline*}
Hence we can rewrite (5.5) of Part I, i.e.,
\begin{equation}\label{det-A}
\det{A} =1 + P_n(a_{j,k})
\end{equation}
where $P_n(a_{j,k})$ is the sum of homogeneous polynomials of
degrees $m=2,\cdots,n$ in the entries $a_{j,k},\;1\leq j <k \leq
n$ with integer coefficients,  as
$$\det A=1+P_n(a_{i,k})=1 \pm 4l^{2}D_{m}^{(-)}(z_{j}-z_{j+1})^{2}+
Q_{n,j}(a_{i,k}).$$ It is clear from (\ref{det-A}),
(\ref{det-details}) and the details provided about the polynomial
$P_n$  that each term of $Q_{n,j}(a_{i,k})$ contains products of
$2-$points functions $D_{m}^{(-)}$ at arguments different from
$z_{j}-z_{j+1}$.
If we choose the arguments $y_{k}^{0}-y_{i}^{0}$
($i < k$) in these $2-$points functions sufficiently large,
$Q_{n,j}(a_{i,k})$ becomes very small; and for these points
$z_{j}$ the determinant $(\det A(z))^{-1/2}$ is holomorphic and
the function $$(\det A(z))^{-1/2}{\mathcal W}^{r}_{0,\alpha
}(z_{1}, \ldots ,z_{n})$$ defines a functional in ${\mathcal
T}(T(L_{j}^{\ell^{\prime}}))^{\prime}$ for any $\ell^{\prime}>\ell
$ by Formula (\ref{eq:n-point}) for all $f \in  {\mathcal T}
(T(L_{j}^{\ell^{\prime}}))$.  In fact, for $\ell ^{\prime}>\ell $,
we choose $\ell^{\prime}> y_{j+1}^{0}-y_{j}^{0}
> \ell $ and other $y_{k}^{0} - y_{i}^{0}$ sufficiently large so
that $(\det A(z))^{-1/2}$ is a bounded function of $x$.  Then the
corresponding integration path $\prod_{j=1}^{n}\Gamma_{j}$ of
(\ref{eq:n-point}) is contained in
$$T(L_{j}^{\ell^{\prime}})=\{z=x+iy\in\C^{4n};
\vert y_{j}-y_{j+1}\vert_{1} < \ell^{\prime}\},$$ where $\vert
y\vert_{1} = \vert y^{0}\vert + \vert \mbox{\bi y}\vert $.  We
conclude that the functional defined by $(\det A(z))^{-1/2}
{\mathcal W} ^{r}_{0,\alpha }(z_{1}, \ldots ,z_{n})$ satisfies
Axiom (i) of (R3).

\vskip 12pt \noindent The transposition of $z_{j}$ and $z_{j+1}$
causes the change of $a_{j,j+1} = a_{j+1,j}$:
$$D_{m}^{(-)}(z_{j}-z_{j+1})\rightarrow D_{m}^{(-)}(z_{j+1}-z_{j})$$
and for an index $k$ with $j < k \neq  j+1$  the change
$$a_{j,k}=a_{k,j}=D_{m}^{(-)}(z_{j}-z_{k})\rightarrow D_{m}^{(-)}
(z_{j+1} - z_{k}) = a_{j+1, k} = a_{k,j+1},$$
$$ a_{j+1,k}=a_{k,j+1}=D_{m}^{(-)}(z_{j+1}-z_{k})\rightarrow
 D_{m}^{(-)}(z_{j} - z_{k}) = a_{j, k} = a_{k,j},$$
results while for an index $k$ with $j > k \neq  j+1$  the change is
$$a_{j,k}=a_{k,j}=D_{m}^{(-)}(z_{k}-z_{j})\rightarrow D_{m}^{(-)}
(z_{k} - z_{j+1}) = a_{j+1,k} = a_{k,j+1},$$
$$a_{j+1,k}=a_{k,j+1}=D_{m}^{(-)}(z_{k}-z_{j+1})\rightarrow
D_{m}^{(-)}(z_{k} - z_{j}) = a_{j, k} = a_{k,j}.$$ We consider the
matrix $B = (b_{i,j})$ obtained from $A$ by the change of $j$-th
and $j+1$-th rows and  $j$-th and $j+1$-th columns. Then we have
$\det A = \det B$.  Next we consider the matrix $C = (c_{j,k})$
obtained from $B$ by changing only $b_{j,j+1} = b_{j+1, j} =
a_{j,j+1} = a_{j+1, j}$, i.e., $c_{j, j+1} = c_{j+1,j} =
D_{m}^{(-)}(z_{j+1} - z_{j})$.  If $x_{j}$ and $x_{j+1}$ are
space-like separated, then $D_{m}^{(-)}(x_{j} - x_{j+1})$ is
analytic (space-like points $x$ are Jost points of
$D_{m}^{(-)}(x)$) and $D_{m}^{(-)}(x_{j} - x_{j+1}) =
D_{m}^{(-)}(x_{j+1} - x_{j})$. Therefore for space-like separated
$x_{j}, x_{j+1}$ ($y_{j}^{0} - y_{j+1}^{0} = 0$) and other
$y_{k}^{0} - y_{i}^{0}$ sufficiently large, we have $\det A = \det
C$.  Note that ${\mathcal W} ^{r}_{0,\alpha }(z_{1}, \ldots
,z_{n})$ is also expressed by the sum of products of the
two-point functions of the Dirac field as in the scalar case, and for
space-like separated $x_{j}, x_{j+1}$ ($y_{j}^{0} - y_{j+1}^{0} =
0$) and other $y_{k}^{0} - y_{i}^{0}$ positive, we have
$${\mathcal W}^{r}_{0,\alpha}(z_{1},\ldots,x_{j},x_{j+1},\ldots,
z_{n})=- {\mathcal W}^{r}_{0,\alpha}(z_{1},\ldots, x_{j+1},x_{j},
\ldots, z_{n}).$$

In order to proceed we need some estimates for
 $D_{m}^{(-)}(x_{j+1} - x_{j})$ which are developed below.

\begin{prop}  Let $\omega (\vert \mbox{\bi p}\vert ) = \sqrt{\vert \mbox{\bi
p}\vert ^{2} + m^{2}}$ and introduce the auxiliary function
$$      g_{m}(z, x) = \int _{0}^{\infty } e^{-i\omega (\vert \mbox{\sbi p}\vert )z}e^{-i\vert \mbox{\sbi p}\vert x} \frac{m}{\vert \mbox{\sbi p}\vert ^{2} + m^{2} + \vert \mbox{\sbi p}\vert \omega (\vert \mbox{\sbi p}\vert )} d\vert \mbox{\bi p}\vert .$$
Then we have
$$      D_{m}^{(-)}(x^{0} - i\epsilon , \mbox{\bi x}) = [(2\pi )^{2}]^{-1} e^{-im(x^{0} - i\epsilon )}\frac{-1}{(x^{0} - i\epsilon )^{2} - \vert \mbox{\sbi x}\vert ^{2}}$$
$$    + \frac{mi}{[2(2\pi )^{2}]} \left[  \frac{-g_{m}(x^{0} - i\epsilon , -\vert \mbox{\bi x}\vert )}{x^{0} - i\epsilon  - \vert \mbox{\sbi x}\vert }       +\frac{g_{m}(x^{0} - i\epsilon , -\vert \mbox{\bi x}\vert )}{x^{0} - i\epsilon  + \vert \mbox{\sbi x}\vert }
\right] ,$$
and for ${\rm Im \, }z \leq  0$ and ${\rm Im \, }x = 0$, the estimate
$$      \vert g_{m}(z, x)\vert  \leq  g_{m}(0, 0) \leq  \frac{\sqrt{2} \pi }{4}$$
follows.
\end{prop}
\begin{proof} From the definition of $D_{m}^{(-)}$ we know
$$      D_{m}^{(-)}(x^{0} - i\epsilon , \mbox{\bi x}) = [2(2\pi )^{3}]^{-1} \int \omega (\vert \mbox{\bi p}\vert )^{-1} e^{-i\omega (\vert \mbox{\sbi p}\vert )(x^{0} - i\epsilon )} e^{i\mbox{\sbi px}} d\mbox{\bi p}$$
$$      = [2(2\pi )^{3}]^{-1} \int \omega (\vert \mbox{\bi p}\vert )^{-1} e^{-i\omega (\vert \mbox{\sbi p}\vert )(x^{0} - i\epsilon )} \exp (i\vert \mbox{\bi p}\vert \vert \mbox{\bi x}\vert  \cos \theta ) \vert \mbox{\bi p}\vert ^{2} \sin \theta  d\vert \mbox{\bi p}\vert  d\theta  d\phi $$
$$      = [2(2\pi )^{2}]^{-1} \frac{1}{i\vert \mbox{\sbi x}\vert } \int _{0}^{\infty }\omega (\vert \mbox{\bi p}\vert )^{-1} e^{-i\omega (\vert \mbox{\sbi p}\vert )(x^{0} - i\epsilon )} [e^{i\vert \mbox{\sbi p}\vert \vert \mbox{\sbi x}\vert } - e^{-i\vert \mbox{\sbi p}\vert \vert \mbox{\sbi x}\vert }] \vert \mbox{\bi p}\vert  d\vert \mbox{\bi p}\vert . $$
If we put $t = \omega (\vert \mbox{\bi p}\vert ) = \sqrt{\vert
\mbox{\bi p}\vert ^{2} + m^{2}}$ then $\vert \mbox{\bi p}\vert
= \sqrt{t^{2} - m^{2}}$, and the equation can be continued by:
$$      = [2(2\pi )^{2}]^{-1} \frac{1}{i\vert \mbox{\sbi x}\vert } \int _{m}^{\infty } e^{-it(x^{0} - i\epsilon )} [e^{i\sqrt{t^{2} - m^{2}}\vert \mbox{\sbi x}\vert } - e^{-i\sqrt{t^{2} - m^{2}}\vert \mbox{\sbi x}\vert }] dt$$
$$      = [2(2\pi )^{2}]^{-1} \frac{1}{i\vert \mbox{\sbi x}\vert } \int _{m}^{\infty } e^{-it(x^{0} - i\epsilon  - \vert \mbox{\sbi x}\vert )} e^{i(\sqrt{t^{2} - m^{2}} - t)\vert \mbox{\sbi x}\vert } dt$$
$$      - [2(2\pi )^{2}]^{-1} \frac{1}{i\vert \mbox{\sbi x}\vert } \int _{m}^{\infty } e^{-it(x^{0} - i\epsilon  + \vert \mbox{\sbi x}\vert )}e^{-i(\sqrt{t^{2} - m^{2}} - t)\vert \mbox{\sbi x}\vert }] dt$$
$$      = [2(2\pi )^{2}]^{-1} \frac{1}{i\vert \mbox{\sbi x}\vert }\left[  \frac{e^{-it(x^{0} - i\epsilon  - \vert \mbox{\sbi x}\vert )} e^{i(\sqrt{t^{2} - m^{2}} - t)\vert \mbox{\sbi x}\vert }}{-i(x^{0} - i\epsilon  - \vert \mbox{\sbi x}\vert )}\right] _{t=m}^{\infty } $$
$$    -  \frac{[2(2\pi)^{2}]^{-1}}{-i(x^{0} - i\epsilon  - \vert \mbox{\sbi x}\vert )} \int _{m}^{\infty } e^{-it(x^{0} - i\epsilon  - \vert \mbox{\sbi x}\vert )} e^{i(\sqrt{t^{2} - m^{2}} - t)\vert \mbox{\sbi x}\vert } \left[ \frac{t}{\sqrt{t^{2} - m^{2}}} - 1\right] dt$$
$$      - [2(2\pi )^{2}]^{-1} \frac{1}{i\vert \mbox{\sbi x}\vert } \left[  \frac{e^{-it(x^{0} - i\epsilon  + \vert \mbox{\sbi x}\vert )} e^{-i(\sqrt{t^{2} - m^{2}} - t)\vert \mbox{\sbi x}\vert }}{-i(x^{0} - i\epsilon  + \vert \mbox{\sbi x}\vert )}\right] _{t=m}^{\infty } $$
$$      +  \frac{[2(2\pi )^{2}]^{-1}}{-i(x^{0} - i\epsilon  + \vert \mbox{\sbi x}\vert )} \int _{m}^{\infty } e^{-it(x^{0} - i\epsilon  + \vert \mbox{\sbi x}\vert )}e^{-i(\sqrt{t^{2} - m^{2}} - t)\vert \mbox{\sbi x}\vert } \left[ \frac{t}{\sqrt{t^{2} - m^{2}}} - 1\right] dt .$$
Since
$$      \int _{m}^{\infty } e^{-it(x^{0} - i\epsilon )}e^{\mp i\sqrt{t^{2} - m^{2}}\vert \mbox{\sbi x}\vert }\left[  \frac{t}{\sqrt{t^{2} - m^{2}}} - 1\right] dt$$
$$      = \int _{0}^{\infty } e^{-i\omega (\vert \mbox{\sbi p}\vert )(x^{0} - i\epsilon )}e^{\mp i\vert \mbox{\sbi p}\vert \vert \mbox{\sbi x}\vert } \left[ 1 - \frac{\vert \mbox{\bi p}\vert }{\omega (\vert \mbox{\sbi p}\vert )}\right] d\vert \mbox{\bi p}\vert $$
$$      = m \int _{0}^{\infty } e^{-i\omega (\vert \mbox{\sbi p}\vert )(x^{0} - i\epsilon )}e^{\mp i\vert \mbox{\sbi p}\vert \vert \mbox{\sbi x}\vert } \frac{m}{\vert \mbox{\sbi p}\vert ^{2} + m^{2} + \vert \mbox{\sbi p}\vert \omega (\vert \mbox{\sbi p}\vert )} d\vert \mbox{\bi p}\vert $$ $$ = m g_{m}(x^{0} - i\epsilon , \pm  \vert \mbox{\bi x}\vert )$$
and
$$      \frac{1}{i\vert \mbox{\sbi x}\vert }\left[  \frac{e^{-it(x^{0} - i\epsilon  - \vert \mbox{\sbi x}\vert )} e^{i(\sqrt{t^{2} - m^{2}} - t)\vert \mbox{\sbi x}\vert }}{-i(x^{0} - i\epsilon  - \vert \mbox{\sbi x}\vert )}\right] _{t=m}^{\infty }$$
$$      - \frac{1}{i\vert \mbox{\sbi x}\vert } \left[  \frac{e^{-it(x^{0} - i\epsilon  + \vert \mbox{\sbi x}\vert )} e^{-i(\sqrt{t^{2} - m^{2}} - t)\vert \mbox{\sbi x}\vert }}{-i(x^{0} - i\epsilon  + \vert \mbox{\sbi x}\vert )}\right] _{t=m}^{\infty }$$
$$      = \frac{1}{i\vert \mbox{\sbi x}\vert }\left[  \frac{-i}{x^{0} - i\epsilon  - \vert \mbox{\sbi x}\vert } + \frac{i}{x^{0} - i\epsilon  + \vert \mbox{\sbi x}\vert } \right] $$
$$      = \frac{-2}{(x^{0} - i\epsilon  - \vert \mbox{\sbi x}\vert )(x^{0} - i\epsilon  + \vert \mbox{\sbi x}\vert )} = \frac{-2}{(x^{0} - i\epsilon )^{2} - \vert \mbox{\sbi x}\vert ^{2}},$$
$$      D_{m}^{(-)}(x^{0} - i\epsilon , \mbox{\bi x}) = [(2\pi )^{2}]^{-1} e^{-im(x^{0} - i\epsilon )}\frac{-1}{(x^{0} - i\epsilon )^{2} - \vert \mbox{\sbi x}\vert ^{2}}$$
$$    + \frac{mi}{[2(2\pi )^{2}]} \left[  \frac{-g_{m}(x^{0} - i\epsilon , -\vert \mbox{\bi x}\vert )}{x^{0} - i\epsilon  - \vert \mbox{\sbi x}\vert } + \frac{g_{m}(x^{0} - i\epsilon, \vert \mbox{\bi x}\vert)}{x^{0} - i\epsilon  + \vert \mbox{\sbi x}\vert}\right] .$$
$$      \vert g_{m}(z, x)\vert  \leq  \int _{0}^{\infty } \frac{m}{\vert \mbox{\sbi p}\vert ^{2} + m^{2} + \vert \mbox{\sbi p}\vert \omega (\vert \mbox{\sbi p}\vert )} d\vert \mbox{\bi p}\vert  = g_{m}(0, 0)$$
$$      \leq  \int _{0}^{\infty } \frac{m}{2\vert \mbox{\sbi p}\vert ^{2} + m^{2}} d\vert \mbox{\bi p}\vert
        = \frac{m}{2} \left[  \frac{\sqrt{2}}{m} \tan ^{-1} \frac{\sqrt{2} \vert \mbox{\bi p}\vert }{m}\right] _{\vert \mbox{\sbi p}\vert  = 0}^{\infty }
        = \frac{\sqrt{2} \pi }{4}.$$
\end{proof}

\begin{cor}  Introduce
$$      \ell _{m}(l) = [1/(2\pi )][l^{2}m\sqrt{2}/8 + l \sqrt{2 + 2(m/8)^{2}l^{2}}] $$
and $a = \min _{\pm } \vert x^{0} - i\epsilon  \pm  \vert
\mbox{\bi x}\vert \vert $.  Then, if $a  > \ell _{m}(l)$, the
estimate
$$      2l^{2}\vert D_{m}^{(-)}(x^{0} - i\epsilon _{\ell }(x), \mbox{\bi x})\vert  < 1$$ holds.
\end{cor}
\begin{proof}  We have the following inequalities.
$$      \vert D_{m}(x^{0} - i\epsilon , \mbox{\bi x})\vert  \leq  (2\pi )^{-2} \left| \frac{1}{(x^{0} - i\epsilon  - \vert \mbox{\sbi x}\vert )(x^{0} - i\epsilon  + \vert \mbox{\sbi x}\vert )} \right|$$
$$    + [2(2\pi )^{2}]^{-1} m \frac{\sqrt{2} \pi }{4} \left[  \left| \frac{1}{x^{0} - i\epsilon  - \vert \mbox{\sbi x}\vert } \right|  + \left| \frac{1}{x^{0} - i\epsilon  + \vert \mbox{\sbi x}\vert } \right|\right] $$
$$      \leq  (2\pi )^{-2}  \frac{1}{a^{2}} + (2\pi )^{-2} m \frac{\sqrt{2} \pi }{4} \frac{1}{a}, $$
$$      2l^{2}\vert D_{m}(x^{0} - i\epsilon , \mbox{\bi x})\vert  \leq  2l^{2}\left[ (2\pi )^{-2} \frac{1}{a^{2}} + (2\pi )^{-2}
m \frac{\sqrt{2} \pi }{4} \frac{1}{a} \right] .$$
As a solution of the inequality
$$      2l^{2}(2\pi )^{-2}\left[ \frac{1}{a^{2}} +  m \frac{\sqrt{2} \pi }{4} \frac{1}{a}\right] < 1,$$
we have
$$      a > \ell _{m}(l) = [1/(2\pi )][l^{2}m\sqrt{2}/8 + l \sqrt{2 + 2(m/8)^{2}l^{2}}].$$
This completes the proof.
\end{proof}

\begin{cor}  Denote by ${\rm dist \, }(x, \bar{V} )$ the distance between $x$ and the closed light cone $\bar{V}  = \{ x = (x^{0}, \mbox{\bi x}) \in
\R ^{4}; \vert x^{0}\vert \geq \vert \mbox{\bi x}\vert \} $, and
for $\ell  > 0$
$$      V_{\ell } = \{ x \in  \R ^{4}; {\rm dist \, }(x, \bar{V} ) < \ell \}.$$
\indent Define $\epsilon _{\ell }(x)$ by $\epsilon _{\ell }(x) =
\ell $ if ${\rm dist \, }(x, \bar{V} ) \leq  \ell /\sqrt{2}$,
$\epsilon _{\ell }(x) = \sqrt{2\ell ^{2} - 2 {\rm dist \, }(x,
\bar{V} )^{2}}$ if $\ell /\sqrt{2} \leq  {\rm dist \, }(x, \bar{V}
) \leq  \ell $ and $\epsilon _{\ell }(x) = 0$ if ${\rm dist \,
}(x, \bar{V} ) \geq \ell $. Then $0 \leq  \epsilon _{\ell }(x)
\leq  \ell $ and ${\rm supp\, }\epsilon _{\ell }(x) \subset
\bar{V} _{\ell }$. Let $\ell  = l/(\sqrt{2}\pi )$ and assume
$\sqrt{2}\ell  > \ell _{m}(l)$, e.g., assume $ml < 2$. Then, if
$\ell ^{\prime \prime }
> \ell $, the estimate
$$      2l^{2}\vert D_{m}^{(-)}(x^{0} - i\epsilon _{\ell ^{\prime \prime }}(x), \mbox{\bi x})\vert  < 1$$ holds.
\end{cor}
\begin{proof}  The support property of $\epsilon _{\ell ^{\prime \prime }}(x)$ follows immediately from the
definitions, and it is easy to see that
$$      \vert x^{0} \pm  \vert \mbox{\bi x}\vert \vert \geq \sqrt{2}\, {\rm dist \, }(x, \bar{V})$$
and we have,
$$      a(x)^{2} = \min _{\pm } \vert x^{0} - i\epsilon _{\ell ^{\prime \prime }}(x) \pm  \vert \mbox{\bi x}\vert \vert ^{2} \geq
2 {\rm dist \, }(x, \bar{V})^{2} + \epsilon _{\ell ^{\prime \prime
}}(x)^{2}.$$ If ${\rm dist \, }(x, \bar{V} ) \geq  \ell ^{\prime
\prime }/\sqrt{2}$, then $a(x)^{2} = 2\ell ^{\prime \prime 2} >
2\ell ^{2} > \ell _{m}(l)^{2}$, and the estimate holds.  If ${\rm
dist \, }(x, \bar{V} ) \leq  \ell ^{\prime \prime }/\sqrt{2}$,
then $\epsilon _{\ell ^{\prime \prime }}(x) = \ell ^{\prime \prime
}
> \ell $, and the estimate follows from the inequality (\ref{estim-2point}).
This completes the proof.
\end{proof}

For any $\ell ^{'}> \ell $, we choose
$\ell<\ell^{\prime\prime}<\ell^{\prime }$.
 \noindent Let $\epsilon(x) = \epsilon _{\ell ^{\prime\prime}}(x)$, and $a_{j,j+1}=D_{m}^{(-)}(x_{j}-x_{j+1}+i\epsilon
(x_{j}-x_{j+1}))$ and for the other $a_{i,k}$ take $y_{k}^{0}-y_{i}^{0}$
sufficiently large. Then $(\det A(x))^{-1/2}$ and $(\det
C(x))^{-1/2}$ are well-defined continuous functions of $x$ and
$(\det A(x))^{-1/2} = (\det C(x))^{-1/2}$ if $x_{j} - x_{j+1} \in
\R^{4} \backslash V_{\ell^{\prime\prime}}$. Let
$${\mathcal W}^{r}_{\alpha}(z_{1},\ldots,z_{n})=(\det A(z))^{-1/2}
{\mathcal W}^{r}_{0,\alpha}(z_{1},\ldots,z_{n})$$ and
$${\mathcal W}^{r,j}_{\alpha}(z)={\mathcal W}^{r^{\prime}}_{\alpha^
{\prime}}(z^{\prime}),z^{\prime}=(z_{1},\ldots,z_{j+1},z_{j},\ldots
, z_{n}),$$
$$r^{\prime}=(r_{1},\ldots,r_{j+1},r_{j},\ldots,_{n}), \alpha^
{\prime}=(\alpha_{1},\ldots,\alpha_{j+1},\alpha_{j},\ldots,\alpha
_{n}).$$ Then, by deforming the path
$\Gamma_{j}\times\Gamma_{j+1}$ in Eq. (\ref{eq:n-point}) into
$G_{j,j+1}$, we can write
\begin{multline*}
{\mathcal W}^{r}_{\alpha}(f)+{\mathcal W}^{r,j}_{\alpha}(f)=\\
\int_{G_{j,j+1} \prod_{i\neq j,j+1}\Gamma_{i}}{\mathcal
W}^{r}_{\alpha}(z)f(z)dz + \int_{G_{j+1,j} \prod_{i\neq
j,j+1}\Gamma_{i}} {\mathcal W}^{r,j}_{\alpha}(z)f(z)
dz,\end{multline*} where $y_{j}^{0} = y_{j+1}^{0}$ and
$$G_{j,j+1}=\{(x_{j}^{0}+ iy_{j}^{0}-i\epsilon (x_{j}-x_{j+1}),
\mbox{\bi x}_{j},x_{j+1}^{0}+iy_{j+1}^{0},\mbox{\bi x}_{j+1});
(x_{j}, x_{j+1}) \in \R^{2\cdot 4}\} ,$$
$$ G_{j+1,j}=\{(x_{j}^{0} +iy_{j}^{0},\mbox{\bi
x}_{j},x_{j+1}^{0}+i
 y_{j+1}^{0} - i\epsilon (x_{j+1}-x_{j}),\mbox{\bi x}_{j+1});(x_{j},
 x_{j+1}) \in\R^{2\cdot 4}\} .$$
Since ${\mathcal W}^{r}_{\alpha}(z)+{\mathcal
W}^{r,j}_{\alpha}(z)=0$ for $x_{j}-x_{j+1}\in\R^{4} \backslash
V^{\ell^{\prime\prime}}$,
\begin{multline*}
{\mathcal W}^{r}_{\alpha}(f)+{\mathcal W}^{r,j}_{\alpha}(f)=\\
\int_{G^{\ell^{\prime}}_{j,j+1} \prod_{i\neq
j,j+1}\Gamma_{i}}{\mathcal W}^{r}_{\alpha}(z)f(z)dz +
\int_{G^{\ell^{\prime}}_{j+1,j} \prod_{i\neq j,j+1}\Gamma_{i}}
{\mathcal W}^{r,j}_{\alpha}(z)f(z) dz,
\end{multline*}
where
\begin{multline*}
G^{\ell^{\prime\prime}}_{j,j+1}=\\ \{(x_{j}^{0}+
iy_{j}^{0}-i\epsilon (x_{j}-x_{j+1}), \mbox{\bi
x}_{j},x_{j+1}^{0}+iy_{j+1}^{0},\mbox{\bi x}_{j+1}); x_{j} -
x_{j+1} \in \R^{4} \cap V^{\ell^{\prime\prime}} \}
,\end{multline*}
\begin{multline*} G^{\ell^{\prime\prime}}_{j+1,j}=\\ \{(x_{j}^{0} +iy_{j}^{0},\mbox{\bi
x}_{j},x_{j+1}^{0}+i
 y_{j+1}^{0} - i\epsilon (x_{j+1}-x_{j}),\mbox{\bi x}_{j+1}); x_{j} -
 x_{j+1} \in\R^{4} \cap V^{\ell^{\prime\prime}} \} .\end{multline*}
Since $G^{\ell^{\prime\prime}}_{j,j+1} \prod_{i\neq
j,j+1}\Gamma_{i}, G^{\ell^{\prime \prime}}_{j+1,j} \prod_{i\neq
j,j+1}\Gamma_{i} \subset W_j ^{\ell ^{\prime}}$, this shows that
$${\mathcal T}(W^{\ell^{\prime}}_{j})\ni f \rightarrow
{\mathcal W}^{r}_{\alpha}(f) + {\mathcal W}^{r,j}_{\alpha }(f)
 \in \C$$ is continuous and satisfies the axiom (ii) of (R3) of
[\cite{BN04}].

\section{Convergence of Wick power series for $\rho(x)= :e^{g\phi(x)^2}:$}
Our starting point are the well-known results of Jaffe [\cite{Ja65}]
on formal Wick power series of free fields. If we consider the power
series of a free field $\phi $
\begin{equation} \label{powerseries}
  \rho ^{(i)}(x) = \sum _{n=0}^{\infty } a^{(i)}_{n}\frac{:\phi  (x)^{n}:}{n!},
\end{equation}
then we have the following theorem.
\begin{thm}[Theorem A.1 of {[\cite{Ja65}]}]
In the sense of  formal power series the following identity holds
\begin{equation} \label{jaffe1}
 (\Phi _{0}, \rho ^{(1)}(x_{1}) \cdots \rho ^{(n)}(x_{n})\Phi _{0})
 = \sum ^{\infty }_{r_{ij}=0;\, 1\leq i<j\leq n} \frac{A(R)T^{R}}{R!}
\end{equation} where
$$   r_{ij} = r_{ji},\; r_{ii} = 0,\; R_{i} = \sum ^{n}_{j=1}
r_{ij},\; A(R) = \prod ^{n}_{j=1} a^{(j)}_{R_{j}}$$
\begin{equation} \label{jaffe2}
  R! = \prod _{1\leq i<j\leq n}(r_{ij})!,\quad T^{R} = \prod _{1\leq i<j\leq n}(t_{ij})^{r_{ij}}
\end{equation}
$$        t_{ij} = (\Phi _{0}, \phi (x_{i})\phi (x_{j})\Phi _{0}) = D^{(-)}_{m}(x_{i}-x_{j}).$$
\end{thm}
\begin{cor} \label{exp}  In the case of
$$      \sigma ^{(i)}(x) = :e^{g_{i}\phi (x)} := \sum _{n=0}^{\infty } g_{i}^{n} \frac{:\phi (x)^{n}:}{n!},$$
(\ref{jaffe1}) becomes
$$      (\Phi _{0}, \sigma ^{(1)}(x_{n}) \cdots \sigma ^{(n)}(x_{n})\Phi _{0}) = \exp \left\{ \sum _{1\leq i<j\leq n} g_{i}g_{j} t_{ij}\right\} .$$
\end{cor}
\begin{proof}
 The chain of identities
$$      \left( \prod _{1\leq i<j\leq n} \{ g_{i}g_{j}\} ^{r_{ij}}\right) ^{2} = \prod _{1\leq i, j\leq n} \{ g_{i}g_{j}\} ^{r_{ij}} = \prod _{i=1}^{n} \prod _{j=1}^{n} g_{i}^{r_{ij}}g_{j}^{r_{ij}}$$
$$      = \prod _{i=1}^{n} \left( g_{i}^{ \sum _{j=1}^{n} r_{ij}} \prod _{j=1}^{n} g_{j}^{r_{ij}}\right) = \left( \prod _{i=1}^{n} g_{i}^{R_{i}}\right)  \prod _{i=1}^{n} \prod _{j=1}^{n} g_{j}^{r_{ij}}$$
$$      = \left( \prod _{i=1}^{n} g_{i}^{R_{i}}\right)  \prod _{j=1}^{n} g_{j}^{\sum _{i=1}^{n} r_{ij}} = \left( \prod _{i=1}^{n} g_{i}^{R_{i}}\right) ^{2} = A(R)^{2}$$
shows that $\displaystyle \prod _{1\leq i<j\leq n} \{ g_{i}g_{j}\}
^{r_{ij}} = A(R)$, and thus we get
$$      \exp \left\{ \sum _{1\leq i<j\leq n} g_{i}g_{j} t_{ij}\right\}  = \prod _{1\leq i<j\leq n} \exp \left\{  g_{i}g_{j} t_{ij}\right\} = \prod _{1\leq i<j\leq n} \sum _{r_{ij}=0}^{\infty } \frac{\{ g_{i}g_{j} t_{ij}\} ^{r_{ij}}}{r_{ij}!}$$
$$      = \sum _{r_{ij}=0; 1\leq i<j\leq n}^{\infty } \prod _{1\leq i<j\leq n}  \frac{\{ g_{i}g_{j} t_{ij}\} ^{r_{ij}}}{r_{ij}!} = \sum _{r_{ij}=0; 1\leq i<j\leq n}^{\infty } \prod _{1\leq i<j\leq n} \{ g_{i}g_{j}\} ^{r_{ij}} \frac{T(R)}{R!}.$$
$$      = \sum _{r_{ij}=0; 1\leq i<j\leq n}^{\infty } \frac{A(R)T(R)}{R!} = (\Phi _{0}, \sigma ^{(1)}(x_{n}) \cdots \sigma ^{(n)}(x_{n})\Phi _{0}).$$
\end{proof}
\indent Assume that for some $\sigma  > 0$
$$      \limsup _{n \rightarrow  \infty } [\vert a^{(i)}_{n}\vert ^{2}/n!]^{1/n} = \sigma .$$
Then Theorem 6.3 of [\cite{BN04}] says that the power series
(\ref{powerseries}) defines an ultra-hyperfunction quantum field with
fundamental length $\ell $
$$      \ell  = \sqrt{\sigma }/(2\pi )$$
if $\phi $ is a massless free field.
Now consider
$$       \rho (x) =  :e^{ig \phi (x)^{2}}: = \sum _{n=0}^{\infty } (ig)^{n}\frac{:\phi (x)^{2n}:}{n!}$$
$$       = \sum _{n=0}^{\infty } (ig)^{n}\frac{(2n)!}{n!} \frac{:\phi (x)^{2n}:}{(2n)!},$$
$$       \rho ^{*}(x) =  :e^{-ig \phi (x)^{2}}: = \sum _{n=0}^{\infty } (-ig)^{n}\frac{:\phi (x)^{2n}:}{n!}.$$
In this case we find for the above limit $\sigma  = 2\vert g\vert $.  Suppose that the
 $0 < t_{ij}$'s satisfy
$$       \sum_{\mathclap{1\leq i<j\leq n}} t_{ij} < \frac{1}{2\vert g\vert }.$$
Then the power series
\begin{equation} \label{jaffe3}
  \sum ^{\infty }_{\mathclap{r_{ij}=0;\, 1\leq i<j\leq n}}\; \frac{A(R)Z^{R}}{R!}
\end{equation}
of $z_{ij}$ $(1\leq i<j\leq n)$  for $\rho ^{(j)}(x) = \rho (x)$ or $\rho ^{(j)}(x) = \rho ^{*}(x)$, where
$Z^{R} = \prod _{1\leq i<j\leq n}(z_{ij})^{r_{ij}}$, is absolutely convergent for $\vert z_{ij}\vert < t_{ij}$
$(1\leq i<j\leq n)$.  This shows the convergence of the vacuum expectation
value
$$       (\Phi _{0}, \rho ^{(1)}(x_{1}) \cdots \rho ^{(n)}(x_{n})\Phi _{0})$$
in the sense of tempered ultra-hyperfunctions, and moreover implies
the strong convergence of
$$      \rho _{N}(f)\Phi  = \sum _{n=0}^{N} (ig)^{n}\frac{:\phi (x)^{2n}:(f)}{n!} \Phi $$
for $N \rightarrow  \infty $ (in the Fock space), where $\Phi  = \rho ^{(1)}(f_{1}) \cdots
\rho ^{(m)}(f_{m})\Phi _{0}$ for $f_{k} \in  {\mathcal T} (T(\mathbb R^{4}))$.
For the definition and basic properties of the testfunction space ${\mathcal T} (T(\mathbb R^{4}))$
of tempered ultrahyperfunctions we refer to [\cite{BN04}].
\begin{prop} \label{prop:vevs}
Abbreviate
$$\rho ^{(j)}(x_{j}) = :e^{-r_{j}il^{2}\phi (x_{j})^{2}}:$$
with $r_{j} = \pm 1$.  Then the vacuum expectation values of these
fields are given by
\begin{equation} \label{2:npoint}
(\Phi _{0} , \rho ^{(1)}(x_{1}) \cdots  \rho ^{(n)}(x_{n})\Phi
_{0} ) = (\det A)^{-1/2},
\end{equation}
where $A$ is the $n \times  n$ symmetric matrix whose entries
$a_{j,k}$ are given by
$$       a_{j, k} = a_{k, j} = 2h_{r_{j}}h_{r_{k}}l^{2}D^{(-)}_{m}(x_{j} - x_{k})$$
for $h_{\pm 1} = e^{\pm i\pi /4}$, $j < k$ and $a_{j, j} = 1$.
\end{prop}
Note that the result (\ref{2:npoint}) is the same as the corresponding result in
[\cite{NB07a}].
\begin{proof}
The equation
$$   (2\pi )^{-1/2} \int e^{itp} e^{-t^{2}/2} dt = e^{-p^{2}/2}$$
can be considered as an equation for the following two power
series of the variable $p$:
$$ (2\pi )^{-1/2} \int  \sum _{n=0}^{\infty }[(itp)^{n}/n!] e^{-t^{2}/2} dt = \sum _{n=0}^{\infty }(-p^{2}/2)^{n}/n! ,$$
and by inserting $p=\sqrt{2}\phi(x)$ and using Wick products we
get, as a formal series
$$(2\pi )^{-1/2} \int  \sum _{n=0}^{\infty }[:(it\sqrt{2}h\phi (x))^{n}:/n!] e^{-t^{2}} dt =
\sum _{n=0}^{\infty }:(-(h\phi (x))^{2})^{n}:/n!.$$ We write this
as
$$(2\pi )^{-1/2} \int :e^{it\sqrt{2}h\phi (x)}: e^{-t^{2}/2} dt = :e^{-(h\phi (x))^{2}}:.$$
Let $h_{r_j}=e^{ir_j \pi/4}$ and denote $\sigma ^{(j)}(x) =
:e^{it_{j}\sqrt{2}h_{r_{j}}\phi (x)}:$.  Then Corollary 2.2 says
$$(\Phi _{0} , \sigma ^{(1)}(x_{1}) \cdots  \sigma ^{(n)}(x_{n})\Phi _{0} )$$
$$= \exp \left\{\sum_{1\leq j<k\leq n} -2t_{j}t_{k}h_{r_{j}}h_{r_{k}}D^{(-)}_{m}
(x_{j} - x_{k})\right\} $$ and thus we get
$$(\Phi_{0}, \rho (x_{1}) \cdots \rho(x_{n})\Phi_{0} ) $$
$$=
\frac{1}{(2\pi)^{n/2}} \int e^{\left\{\sum_{{1\leq j<k\leq n}}
-2t_{j}t_{k}h_{r_{j}}h_{r_{k}}D^{(-)}_{m}(x_{j} - x_{k}) -
\sum_{j=1}^{n} t_{j}^{2}/2\right\} } dt_{1} \ldots dt_{n}$$
$$       = (\det A)^{-1/2}.$$
\end{proof}
\indent Note that $\phi (x) = \rho ^{(i)}(x)$ if $a^{(i)}_{1} = 1$
and $a^{(i)}_{n} = 0$ for $n \neq  1$.  Let $U(a, \Lambda )$ be
the unitary representation of the proper Poincar\'e group for the
free neutral scalar field in the Fock space $ {\mathcal H} $.
Then the system $\{  {\mathcal H} , \Phi _{0}, U(a, \Lambda ),
\phi (x), \rho (x), \rho ^{*}(x)\} $ satisfies the axioms of
UHFQFT. \vskip 12pt \noindent

\section{Verification of the equation $\partial_{\mu}\rho(x)=2i\ell^2 :\rho(x)\phi(x)\partial_{\mu}\phi(x):$} \noindent
We begin by recalling some basic facts about Wick products of free
fields which are then used to study  Wick polynomials and Wick power
series.  \vskip 12pt \noindent
Let ${\mathcal H} $ be the Hilbert space defined by $$
{\mathcal H}  = \oplus _{n=0}^{\infty } {\mathcal H} _{n}.$$  Here, ${\mathcal H} _{n}$ is
the set of symmetric square-integrable functions on the direct
product of the momentum space hyperboloids \beq
\label{eq:hyperboloid}    \xi _{k}^{2} = m^{2}, \  \xi _{k}^{0} > 0, \  k = 1, \ldots , n
\eeq with respect to the Lorentz invariant measure $\displaystyle \prod
_{k=1}^{n} d\Omega _{m}(\xi _{k})$, where $$        d\Omega _{m}(\xi ) = \frac{d\xi ^{1} d\xi ^{2} d\xi ^{3}}{\sqrt{
\sum _{k=1}^{3} (\xi ^{k})^{2} + m^{2}}}.$$
In the fundamental paper [\cite{WG64}], we find the following quite
general formula (3.44) for the definition of Wick products of a free
field $\phi $ of mass $m$ as operators in $\h$: For $f \in  {\mathcal S} (
\mathbb R^{4})$ and $\Phi  \in  {\mathcal H} $ one has:
$$ (:D^{\alpha^{(1)}}\phi D^{\alpha^{(2)}}\phi \cdots
D^{\alpha^{(l)}}\phi : (f) \Phi)^{(n)} (\xi_{1}, \ldots , \xi _{n})
\eqno{(3.44)}$$
$$ = \frac{\pi^{l/2}}{(2\pi)^{2(l-1)}} \sum_{j=0}^{l} \left[
\frac{(n-l+2j)!}{n!} \right]^{1/2} \int \cdots \int \left(
\prod_{k=1}^{j} d\Omega _{m}(\eta _{k})\right)\times  $$
$$\sum_{\mathclap{1 \leq k_{1}<k_{2}< \ldots < k_{l-j} \leq n}}\;\;
{(j!)^{-1}} \sum_{P} P \left((-i\eta_{1})^{\alpha^{(1)}} \cdots
(-i\eta_{j})^{\alpha^{(j)}}(i\xi_{k_{1}})^{\alpha^{(j+1)}}\cdots
\right.$$
$$ \cdots (i\xi_{k_{l-j}})^{\alpha^{(l)}}\left. \tilde{f}\left(\sum_{r=1}^{j} \eta_{r} -
\sum_{r=1}^{l-j}\xi_{k_{r}}\right)\right)$$
$$\Phi^{(n-l+2j)}(\eta_{1},
\ldots, \eta_{j}, \xi_{1}, \ldots, \hat{\xi}_{k_{1}},\ldots,
\hat{\xi}_{k_{l-j}},\ldots,\xi_{n}),$$
where in the summation $\displaystyle \sum _{j=0}^{l}$, only those terms are
to be retained for which $n - l + 2j \geq  0$, and the sum
$\displaystyle \sum _{P}$ is over all permutation of the variables $\eta _{1},
\ldots , \eta _{j}, (-\xi _{k_{1}}), \ldots , (-\xi _{k_{l-j}})$.
We reconsider this formula in the sense of operator-valued ultra-hyperfunctions.
Let $\vert \beta \vert  = 1$ and $\vert \alpha  ^{(1)}\vert  = \vert \alpha ^{(2)}\vert = \ldots = \vert \alpha  ^{(l)} \vert  = 0$.  Then
we have from (3.44)
$$(:\phi^{l}:(-D^{\beta}f)\Phi)^{(n)}(\xi_{1},\ldots,\xi_{n})
$$
$$
= \frac{\pi^{l/2}}{(2\pi)^{2(l-1)}} \sum_{j=0}^{l} \left[
 \frac{(n-l+2j)!}{n!} \right] ^{1/2} \int \cdots \int \left(
 \prod _{k=1}^{j} d\Omega _{m}(\eta _{k})\right)
 $$
$$
\times\sum_{\mathclap{1\leq k_{1}<k_{2}< \ldots < k_{l-j} \leq n}}
(j!)^{-1}
$$
$$
\times \sum_{P}P\left(i\left(\sum_{r=1}^{j} \eta_{r}-
\sum_{r=1}^{l-j} \xi_{k_{r}} \right)^{\beta} \tilde{f}\left(
\sum_{r=1}^{j}\eta_{r}-\sum_{r=1}^{l-j}\xi_{k_{r}} \right) \right)\times
$$
$$
\times
\Phi^{(n-l+2j)}(\eta_{1},\ldots, \eta_{j}, \xi _{1}, \ldots ,
\hat{\xi}_{k_{1}}, \ldots, \hat{\xi}_{k_{l-j}}, \ldots, \xi_{n}).$$
$$= \frac{\p ^{l/2}}{(2\pi)^{2(l-1)}} \sum_{j=0}^{l}\left[\frac{(n-l+2j)!}{n!}
\right]^{1/2}\int \cdots \int \left( \prod _{k=1}^{j} d\Omega _{m}(\eta _{k})\right) $$
$$\times\sum_{\mathclap{1\leq k_{1}<k_{2}< \cdots < k_{l-j} \leq  n}} (j!)^{-1} \sum_{P} P
\left( l(i\eta_{1})^{\beta} \tilde{f} \left(\sum_{r=1}^{j}\eta_{r}-
\sum_{r=1}^{l-j} \xi_{k_{r}} \right) \right)
 $$
$$
 \times\Phi ^{(n-l+2j)}(\eta _{1}, \ldots , \eta _{j}, \xi _{1}, \ldots , \hat{\xi }   _{k_{1}},
\ldots , \hat{\xi }   _{k_{l-j}}, \ldots , \xi _{n}).$$
Observe that
$$        \sum _{P} P\left( \eta _{i} \right) = \sum _{P} P \left( -\xi _{k_{r}}\right) $$
for any $i$ and $r$.  This implies for $\vert \beta \vert = 1$,
$$        \sum _{P} P\left( (\eta _{i})^{\beta } \right) = \sum _{P} P \left( (-\xi  _{k_{r}})^{\beta } \right) $$
and therefore
$$    \sum _{P} P\left( i\left( \sum _{r=1}^{j} \eta _{r} - \sum _{r=1}^{l-j} \xi  _{k_{r}} \right) ^{\beta }  \tilde{f}   \left( \sum _{r=1}^{j} \eta _{r} - \sum _{r=1}^{l-j}\xi _{k_{r}} \right) \right) $$
$$      =  \sum _{P} P\left( l(i\eta _{1})^{\beta } \tilde{f}   \left( \sum _{r=1}^{j} \eta _{r} - \sum _{r=1}^{l-j} \xi _{k_{r}} \right) \right)  $$
On the other hand, we also have  from (3.44), for $\vert \alpha ^{(1)}\vert  = 1$, and
$\vert \alpha ^{(2)}\vert  = \ldots = \vert \alpha ^{(l)} \vert  = 0$
$$        (:(D^{\alpha ^{(1)}}\phi ) \phi ^{l-1}: (f) \Phi )^{(n)} (\xi _{1}, \ldots , \xi _{n})$$
$$        = \frac{\pi ^{l/2}}{(2\pi )^{2(l-1)}} \sum _{j=0}^{l} \left[  \frac{(n-l+2j)!}{n!}\right] ^{1/2}
\int \cdots \int \left( \prod _{k=1}^{j} d\Omega _{m}(\eta _{k})\right) $$
$$\times\sum_{\mathclap{1 \leq   k_{1}<k_{2}< \ldots < k_{l-j} \leq n}}\; (j!)^{-1} \sum _{P} P
\left( (i\eta _{1})^{\alpha ^{(1)}} \tilde{f}   \left( \sum _{r=1}^{j} \eta _{r} - \sum _{r=1}^{l-j}  \xi _{k_{r}} \right) \right) $$
$$        \times  \Phi ^{(n-l+2j)}(\eta _{1}, \ldots , \eta _{j}, \xi _{1}, \ldots , \hat{\xi }   _{k_{1}},
\ldots , \hat{\xi }   _{k_{l-j}}, \ldots , \xi _{n}).$$
This shows that
\beq \label{eq:der1}  (:\phi ^{l}: (-D^{\alpha ^{(1)}}f)\Phi )^{(n)} = l (:(D^{\alpha ^{(1)}}\phi ) \phi ^{l-1}:(f)\Phi )^{(n)},\eeq
that is,
\beq \label{eq:der2}
   D^{\alpha ^{(1)}}:\phi (x)^{l}:= l :(D^{\alpha ^{(1)}}\phi (x)) \phi ^{l-1}(x):.\eeq
Let ${\mathcal D} _{0}$ be the set generated by the vectors of the form
$$      \rho ^{(1)}(f_{1}) \cdots \rho ^{(n)}(f_{n})\Phi _{0}, f_{k} \in  {\mathcal T} (T(\mathbb R^{4})),$$
where $\rho ^{(k)}(x)$ is one of $\phi (x)$, $\rho (x)$ and $\rho ^{*}(x)$, and $\Phi  \in  {\mathcal D} _{0}$.
Then we have seen in the previous section that
$$\rho (-D^{\alpha^{(1)}}f)\Phi = :e^{ig \phi^{2}}:(-D^{\alpha^{(1)}}f)\Phi $$
$$ =\sum_{l=0}^{\infty} \frac{(ig)^{l}}{l!}:\phi ^{2l}: (-D^{\alpha ^{(1)}}f) \Phi $$
is strongly convergent, and by (\ref{eq:der1})
$$ :\phi^{2l}: (-D^{\alpha^{(1)}}f) \Phi = l :(D^{\alpha^{(1)}}\phi) \phi^{l-1}:(f)
\Phi .$$
This shows that
$$\sum_{l=0}^{\infty}\frac{(ig)^{l}}{l!}:\phi^{2l}:(-D^{\alpha^{(1)}}f)\Phi$$
$$ = \sum_{l=1}^{\infty}\frac{(ig)^{l}}{(l-1)!} 2 :(D^{\alpha ^{(1)}}\phi ) \phi  \phi ^{2(l-1)}:
(f)\Phi$$
$$=\sum_{l=0}^{\infty} 2(ig)\frac{(ig)^{l}}{l!} :(D^{\alpha^{(1)}}\phi)\phi
\phi^{2l}:(f)\Phi.$$
We write the last expression as
$$=2(ig):(D^{\alpha^{(1)}}\phi)\phi \sum_{l=0}^{\infty}\frac{(ig)^{l}}{l!}: \phi
^{2l}::
(f)\Phi$$
$$=2ig :(D^{\alpha^{(1)}}\phi) \phi \rho:(f)\Phi .$$
That is, the formal expression (which is difficult to give a direct meaning)
$$ 2ig:(D^{\alpha^{(1)}}\phi(x))\phi(x)(:e^{ig \phi(x)^{2}}:) : \Phi $$
$$= 2ig :(D^{\alpha^{(1)}}\phi(x)) \phi(x)\sum_{l=0}^{\infty}\frac{(ig)^{l}}{l!}
 :\phi ^{2l}(x): : \Phi  $$
can be understood as
$$\sum_{l=0}^{\infty} 2ig :(D^{\alpha^{(1)}}\phi(x))\phi(x)\frac{(ig)^{l}}{l!}\phi ^{2l}(x):\Phi $$
$$=\sum_{l=1}^{\infty} 2 :(D^{\alpha^{(1)}}\phi(x))\frac{(ig)^{l}}{(l-1)!} \phi^{2l-1}(x): \Phi .$$
Then by (\ref{eq:der2}), the above expression equals
$$\sum_{l=1}^{\infty}\frac{(ig)^{l}}{l!} D^{\alpha^{(1)}}:\phi^{2l}(x):\Phi , $$
and this is equal to
$$D^{\alpha^{(1)}} \sum_{l=1}^{\infty} \frac{(ig)^{l}}{l!} : \phi ^{2l}(x):  \Phi  = D^{\alpha ^{(1)}} \rho (x) \Phi $$
in the sense of generalized functions.  In the above understanding,
we have
\beq \label{eq:deriv-field}
D^{\alpha^{(1)}}\rho(x)\Phi=2ig:(D^{\alpha^{(1)}}\phi(x))\phi(x)\rho(x):\Phi,\eeq
that is, if the Wick product
$$ :(D^{\alpha^{(1)}}\phi(x)) \phi(x)\rho(x) :$$
is defined by the Wick power series
$$\sum_{l=0}^{\infty} 2ig :(D^{\alpha^{(1)}}\phi(x))\phi(x)\frac{(ig)^{l}}{l!}
 \phi^{2l}(x):,$$
then we have (\ref{eq:deriv-field}), i.e., (\ref{eq:Wpsder}).
\vskip 12pt \noindent

\section{Wightman's Axioms for general type fields} \label{waxioms}
In Wightman's scheme, the concept of a relativistic quantum field
$\phi ^{(\kappa )}$ of type $\kappa $ plays a fundamental role.  Such a field, for
example a scalar, tensor or spinor field, has a finite number of
Lorentz components $\phi ^{(\kappa )}_{j}$ $(j = 1, \ldots , r_{\kappa })$.
\vskip 12pt \noindent
The field components $\phi ^{(\kappa )}_{j}(x)$ are operator-valued generalized
functions, i.e.,
$$        \phi ^{(\kappa )}_{j}(f) = \int \phi ^{(\kappa )}_{j}(x)f(x) d^{4} x$$
are densely defined linear operators in a complex Hilbert space
${\mathcal H} $.  They are not assumed to be bounded. \vskip 12pt
\noindent Here we state Wightman's axioms for the
ultra-hyperfunction quantum field theory [\cite{BN04}].  For the
neutral scalar fields, these axioms are the axioms in
[\cite{BN04}].
\vskip 12pt \noindent W.I. {\bf Relativistic
invariance of the state space}: There is a complex Hilbert space
${\mathcal H} $ with positive metric in which a unitary
representation $U(a, A)$ of the Poinar\'e spinor group ${\mathcal
P} _{0}$ acts. $(a,A) \mapsto U(a,A)$ is weakly continuous. \vskip
12pt \noindent W.II. {\bf Spectral property}: The spectrum $\Sigma
$ of the energy-momentum operator $P$ which generates the
translations in this representation, i.e., $e^{iaP} = U(a, 1)$, is
contained in the closed forward light cone\newline
 \indent $\bar{V}  _{+} = \{ p = (p^{0}, \ldots, p^{3}) \in  \mathbb R^{4}; p^{0} \geq  \vert  \mbox{\bi p}\vert \} $.
\vskip 12pt \noindent
W.III. {\bf Existence and uniqueness of the vacuum}:  In
${\mathcal H} $ there exists unit vector $\Phi _{0}$ (also denoted by $
\left. \vert 0 \right>$ and called the vacuum vector) which is unique up
to a phase factor and  which is invariant under all  space-time
translations $U(a, 1)$, $a \in  \mathbb R^{4}$.
\vskip 12pt \noindent
W.IV. {\bf Fields}: The components $\phi ^{(\kappa )}_{j}$ of the quantum field
$\phi ^{(\kappa )}$ are operator-valued generalized functions $\phi ^{(\kappa )}_{j}(x)$ over the
space ${\mathcal T} (T(\mathbb R^{4}))$ with common dense domain  ${\mathcal D} $;
 i.e., for all $\Psi \in  {\mathcal D} $ and all $\Phi \in  {\mathcal H} $,
$$        {\mathcal T} (T(\mathbb R^{4})) \ni  f \rightarrow  (\Phi , \phi ^{(\kappa )}_{j}(f)\Psi )  \in  \mathbb C$$
is a tempered ultrahyperfunction.  It is supposed that the vacuum vector
$\Phi _{0}$ is contained in ${\mathcal D} $ and that ${\mathcal D} $ is taken into itself under
the action of the operators $\phi ^{(\kappa )}_{j}(f)$ and $U(a, A)$, i.e.,
$$        \phi ^{(\kappa )}_{j}(f){\mathcal D}  \subset  {\mathcal D} , \  U(a, A){\mathcal D} \subset  {\mathcal D} .$$
Moreover it is assumed that there exist indices $\bar{\kappa }  $, $\bar{\jmath } $
such that $\phi ^{(\bar{\kappa }  )}_{\bar{\jmath }}(\bar{f}   ) \subset  \phi ^{(\kappa )}_{j}(f)^{*}$ where $^{*}$
indicates the Hilbert space adjoint of the operator in question.
\vskip 12pt \noindent
W.V. {\bf Poincar\'e-covariance of the fields}: According to
the type of the field, there is a finite dimensional real or complex
matrix representation $V^{(\kappa )}(A)$ of $SL(2,\mathbb C)$ such that
$$        U(a, A)\phi ^{(\kappa )}_{j}(x)U(a, A)^{-1} = \sum _{\ell } V^{(\kappa )}_{j,\ell }(A^{-1}) \phi ^{(\kappa )}_{\ell }(\Lambda (A)x + a),$$
i.e., for any $f \in  {\mathcal T} (T(\mathbb R^{4}))$ and $\Psi  \in  {\mathcal D} $,
$$        U(a, A)\phi ^{(\kappa )}_{j}(f)U(a, A)^{-1}\Psi = \sum _{\ell } V^{(\kappa )}_{j,\ell }(A^{-1})
 \phi ^{(\kappa )}_{\ell }(f_{(a, A)})\Psi ,$$
where $f_{(a, A)}(x) = f(\Lambda (A)^{-1}(x - a))$. We have $V^{(\kappa )}(-1) = \pm 1$.  If
$V^{(\kappa )}(-1) = 1$, then the field is called a tensor field.  If
$V^{(\kappa )}(-1) = -1$, then the field is called a spinor field.
\vskip 12pt \noindent
W.VI. {\bf Extended causality or extended local commutativity}:
Any two field components $\phi ^{(\kappa )}_{j}(x)$ and $\phi ^{(\kappa ^{\prime })}_{l}(y)$ either commute or
anti-commute if the distance between $x$ and $y$ is greater than $\ell $:\newline
a) The functionals
$$
 {\mathcal T} (T(\mathbb R^{4}))\times  {\mathcal T} (T(\mathbb R^{4})) \ni  f \otimes  g \rightarrow  (\Phi , \phi ^{(\kappa )}_{j}(f)\phi ^{(\kappa ^{\prime })}_{l}(g)\Psi )
$$
and\newline
$$
 {\mathcal T} (T(\mathbb R^{4}))\times  {\mathcal T} (T(\mathbb R^{4})) \ni  f \otimes  g \rightarrow  (\Phi , \phi ^{(\kappa ^{\prime })}_{l}(g)\phi ^{(\kappa )}_{j}(f)\Psi )
$$
can be extended continuously to ${\mathcal T} (T(L^{\ell }))$ in some Lorentz frame,
for arbitrary elements $\Phi $, $\Psi $ in the common domain ${\mathcal D} $ of the
field operators $\phi ^{\kappa }_{j}(f)$, where
$$      T(L^{\ell }) = \{ (z_{1}, z_{2}) \in  \mathbb C^{4\cdot 2}; \vert {\rm Im \, }z_{1} - {\rm Im \, }z_{2}\vert  < \ell \} .$$
b) The carrier of the functional
$$
 (f,  g) \rightarrow  (\Phi , [\phi ^{(\kappa )}_{j}(f), \phi ^{(\kappa ^{\prime })}_{l}(g)]_{\mp }\Psi )
$$
on ${\mathcal T} (T(\mathbb R^{4}))\times  {\mathcal T} (T(\mathbb R^{4}))$ is contained in the set
$$        W^{\ell } = \{ (z_{1}, z_{2}) \in  \mathbb C^{4\cdot 2};  z_{1} - z_{2} \in   V^{\ell }\} ,$$
where
$$  V^{\ell } = \{ z \in  \mathbb C^{4}; \exists \,  x \in  V, \vert {\rm Re \, }z - x \vert   + \vert {\rm Im \, }z \vert < \ell \} $$
is a complex neighborhood of light cone $V$, i.e., this  functional
can be extended continuously to ${\mathcal T} (W^{\ell })$.
\vskip 12pt \noindent
W.VII. {\bf Cyclicity of the vacuum}: The set ${\mathcal D} _{0}$ of
finite linear combinations of vectors of the form
$$        \phi ^{(\kappa _{1})}_{j_{1}}(f_{1}) \cdots \phi ^{(\kappa _{n})}_{j_{n}}(f_{n}) \Phi _{0}, \  f_{j} \in  {\mathcal T} (T(\mathbb R^{4})) \
(n = 0, 1, \ldots )$$
is dense in ${\mathcal H} $.
\vskip 12pt \noindent

\section{Some Consequences of the Axioms}
A vector-valued generalized function $\Phi ^{(\underline{\kappa }_{n})}_{\underline{\mu }_{n}}(f)$ is defined as
follows: First, let $g(x_{1}, \ldots , x_{n}) = f_{1}(x_{1}) \cdots f_{n}(x_{n})$ for
$f_{j} \in  {\mathcal T} (T(\mathbb R^{4}))$, and define $\Phi ^{(\kappa _{1} \ldots \kappa _{n})}_{\mu _{1} \ldots \mu _{n}}(g)$ by:
$$      \Phi ^{(\kappa _{1} \ldots \kappa _{n})}_{\mu _{1} \ldots \mu _{n}}(g) = \phi ^{(\kappa _{1})}_{\mu _{1}}(f_{1}) \cdots \phi ^{(\kappa _{j})}_{\mu _{j}}(f_{j}) \cdots \phi ^{(\kappa _{n})}_{\mu _{n}}(f_{n})\Phi _{0}.$$
If ${\mathcal T} (T(\mathbb R^{4}))^{\otimes n} \ni g_{k}
\rightarrow f(x_{1}, \ldots , x_{n}) \in {\mathcal T} (T(\mathbb
R^{4n}))$ in the topology of ${\mathcal T} (T(\mathbb R^{4n}))$,
$$      \Vert \Phi ^{(\kappa _{1} \ldots \kappa _{n})}_{\mu _{1} \ldots \mu _{n}}(g_{k} - g_{l})\Vert ^{2}$$
$$      = {\mathcal W} ^{(\bar{\kappa } _{n} \ldots \bar{\kappa } _{1} \kappa _{1} \ldots \kappa _{n})}_{\bar{\jmath} _{n} \ldots \bar{\jmath} _{1} j_{1} \ldots j_{n}}((g_{k} - g_{l})^{*} \otimes (g_{k} - g_{l})) \rightarrow  0.$$
This shows that there exists a vector $\Phi ^{(\kappa _{1} \ldots \kappa _{n})}_{\mu _{1} \ldots \mu _{n}}(f)$ such that
$$      \Phi ^{(\kappa _{1} \ldots \kappa _{n})}_{\mu _{1} \ldots \mu _{n}}(g_{k}) \rightarrow  \Phi ^{(\kappa _{1} \ldots \kappa _{n})}_{\mu _{1} \ldots \mu _{n}}(f) = \Phi ^{(\underline{\kappa }_{n})}_{\underline{\mu }_{n}}(f),$$
and the mapping
$$      {\mathcal T} (T(\mathbb R^{4n})) \ni  f \rightarrow  \Phi ^{(\kappa _{1} \ldots \kappa _{n})}_{\mu _{1} \ldots \mu _{n}}(f) \in  {\mathcal H} $$
is continuous.  The Wightman (generalized) function $
{\mathcal W} ^{(\kappa _{1} \ldots \kappa _{n})}_{\mu _{1} \ldots \mu _{n}}(f)$ is defined by
$$      {\mathcal T} (T(\mathbb R^{4n})) \ni  f \rightarrow  {\mathcal W} ^{(\kappa _{1} \ldots \kappa _{n})}_{\mu _{1} \ldots \mu _{n}}(f) = (\Phi _{0}, \Phi ^{(\kappa _{1} \ldots \kappa _{n})}_{\mu _{1} \ldots \mu _{n}}(f)) \in  \mathbb C.$$
With the definition of the Fourier transform $\tilde{\Phi } ^{(\underline{\kappa }_{n})}_{\underline{\mu }_{n}}$ of
$\Phi ^{(\underline{\kappa }_{n})}_{\underline{\mu }_{n}}$ by
$$      \Phi ^{(\underline{\kappa }_{n})}_{\underline{\mu }_{n}}(f) = \tilde{\Phi } ^{(\underline{\kappa }_{n})}_{(\underline{\mu }_{n})}(\tilde{f} ).$$
we find
\begin{multline*}
 U(a,1) \tilde{\Phi}^{(\underline{\kappa }_{n})}_{\underline{\mu }_{n}}
(\tilde{f}) = U(a,1) \Phi^{(\underline{\kappa }_{n})}_{\underline{\mu }_{n}}(f) =\\
 \Phi^{(\underline{\kappa }_{n})}_{\underline{\mu }_{n}}(f_{(a,1)}) =
 \tilde{\Phi} ^{(\underline{\kappa }_{n})}_{\underline{\mu }_{n}}\left( \tilde{f}
   \exp \left[ i\left( \sum _{k=1}^{n} p_{k} a \right) \right] \right) .\end{multline*}
\indent According to standard strategy we use this identity to determine support properties
of the Fourier transforms of the field operators.
Let $h \in  {\mathcal T} (T(\mathbb R^{4}))$.  Then we have
$$      (2\pi )^{2} \tilde{h} (P)\tilde{\Phi } ^{(\underline{\kappa }_{n})}_{\underline{\mu }_{n}}(\tilde{f} ) = \int _{ \mathbb R^{4}} h(a) U(a, 1) da \tilde{\Phi } ^{(\underline{\kappa }_{n})}_{\underline{\mu }_{n}}(\tilde{f} )$$
$$      = (2\pi )^{2} \langle \tilde{\Phi } ^{(\underline{\kappa }_{n})}_{\underline{\mu }_{n}}(p_{1}, \ldots , p_{n}), \tilde{h} (p_{1}+ \cdots + p_{n})\cdot \tilde{f} (p_{1}, \ldots , p_{n})\rangle .$$
Let $\chi _{n}$ be a linear mapping defined by
$$      (p_{1}, \ldots , p_{n}) = \chi _{n}(q_{0}, \ldots , q_{n-1}), \  p_{k} = q_{k-1} - q_{k} (k = 1, \ldots , n-1), \  p_{n} = q_{n-1}.$$
The inverse mapping $\chi _{n}^{-1}$ is:
$$      q_{k} = \sum _{j=k+1}^{n} p_{j} \quad  (k = 0, \ldots , n-1).$$
Define $\tilde{Z}
^{(\underline{\kappa }_{n})}_{\underline{\mu }_{n}}$ by
$$      \tilde{Z} ^{(\underline{\kappa }_{n})}_{\underline{\mu }_{n}}(\tilde{f} \circ \chi _{n}) = \tilde{\Phi } ^{(\underline{\kappa }_{n})}_{\underline{\mu }_{n}}(\tilde{f} ).$$
Then
$$      \tilde{Z} ^{(\underline{\kappa }_{n})}_{\underline{\mu }_{n}}(\tilde{g} ) = \tilde{\Phi } ^{(\underline{\kappa }_{n})}_{\underline{\mu }_{n}}(\tilde{g} \circ \chi _{n}^{-1}).$$
Let $\tilde{g} _{2} \in  H(\mathbb R^{4(n-1)}; \mathbb R^{4(n-1)})$ and $\tilde{g} _{1} \in  H(\mathbb R^{4}; \mathbb R^{4})$.  Then we have
$$      \tilde{h} (P) \tilde{Z} ^{(\underline{\kappa }_{n})}_{\underline{\mu }_{n}}(\tilde{g}  _{1} \otimes \tilde{g} _{2})) = \tilde{Z} ^{(\underline{\kappa }_{n})}_{\underline{\mu }_{n}}(\tilde{h} \cdot \tilde{g}  _{1} \otimes \tilde{g} _{2}))$$
$$      = \tilde{g}_{1} (P) \tilde{Z} ^{(\underline{\kappa }_{n})}_{\underline{\mu }_{n}}(\tilde{h} \otimes \tilde{g} _{2})).$$
These equalities show that the vector-valued generalized function
$$       H(\mathbb R^{4}; \mathbb R^{4}) \ni  \tilde{g} _{1} \rightarrow  \tilde{Z} ^{(\underline{\kappa }_{n})}_{\underline{\mu }_{n}}(\tilde{g}  _{1} \otimes \tilde{g} _{2})) \in  {\mathcal H} $$
has its support contained in the spectrum $\Sigma $ of
energy-momentum operator $P$ (see Proposition 4.5 of [\cite{BN04}]),
and
$$      \tilde{h} (0) (\Phi _{0}, \tilde{Z} ^{(\underline{\kappa }_{n})}_{\underline{\mu }_{n}}(\tilde{g}  _{1} \otimes \tilde{g} _{2}))) = (\Phi _{0}, \tilde{h} (P) \tilde{Z} ^{(\underline{\kappa }_{n})}_{\underline{\mu }_{n}}(\tilde{g}  _{1} \otimes \tilde{g} _{2})))$$
$$      = (\Phi _{0}, \tilde{g} _{1}(P) \tilde{Z} ^{(\underline{\kappa }_{n})}_{\underline{\mu }_{n}}(\tilde{h} \otimes \tilde{g} _{2})) = \tilde{g} _{1}(0) (\Phi _{0}, \tilde{Z} ^{(\underline{\kappa }_{n})}_{\underline{\mu }_{n}}(\tilde{h} \otimes \tilde{g} _{2})).$$
This equality allows us to define a functional $\tilde{W} ^{(\underline{\kappa }_{n})}_{\underline{\mu }_{n}}$ by
$$      (2\pi)^{2}\tilde{W} ^{(\underline{\kappa }_{n})}_{\underline{\mu }_{n}}(\tilde{g} _{2})
 = (\Phi _{0}, \tilde{Z} ^{(\underline{\kappa }_{n})}_{\underline{\mu }_{n}}(\tilde{h} \otimes \tilde{g} _{2}))$$
for $\tilde{h} \in  H(\mathbb R^{4}; \mathbb R^{4})$ with
$\tilde{h}(0) = 1$, since the right hand side of the above equality
does not depend on $\tilde{h} \in  H(\mathbb R^{4}; \mathbb
R^{4})$ provided $\tilde{h}(0) = 1$, equivalently, $\displaystyle
\int h(x) dx = (2\pi )^{2}$.  Moreover, we have
$$      (2\pi )^{2} \tilde{g} _{1}(0) \tilde{W} ^{(\underline{\kappa }_{n})}_{\underline{\mu }_{n}}(\tilde{g} _{2}) = (\Phi _{0}, \tilde{Z} ^{(\underline{\kappa }_{n})}_{\underline{\mu }_{n}}(\tilde{g} _{1} \otimes \tilde{g} _{2})),$$
and this shows that
$$      \tilde{{\mathcal W}}^{(\underline{\kappa }_{n})}_{\underline{\mu }_{n}}\circ \chi _{n}(q_{0}, q_{1}, \ldots , q_{n-1}) = (2\pi )^{2} \delta (q_{0}) \tilde{W} ^{(\underline{\kappa }_{n})}_{\underline{\mu }_{n}}(q_{1}, \ldots , q_{n-1}).$$
Let $\tilde{f} _{j} = \tilde{g} _{j}\circ \chi _{n}^{-1} \  (j = 1, 2)$.  Then
$$      f^{*}_{1} \otimes f_{2}^{~}(p_{1}, \ldots , p_{m+n}) = \overline{\tilde{f} }_{1}(-p_{m}, \ldots , -p_{1}) f_{2}^{~}(p_{m+1}, \ldots , p_{m+n})$$
$$      = \overline{\tilde{g} }_{1}(-p_{1} - \cdots - p_{m}, \ldots , -p_{1}) \tilde{g} _{2}(p_{m+1} + \cdots + p_{m+n}, \ldots , p_{m+n})$$
$$      = \overline{\tilde{g} }_{1}(q_{m} - q_{0}, \ldots , q_{1} - q_{0}) \tilde{g} _{2}(q_{m}, \ldots , q_{m+n-1}),$$
and
$$      (\tilde{Z} ^{(\underline{\kappa }_{m})}_{\underline{\mu }_{m}}(\tilde{g} _{1}), \tilde{Z} ^{(\underline{\kappa }_{n})}_{\underline{\mu }_{n}}(\tilde{g} _{2})) = (\tilde{\Phi } ^{(\underline{\kappa }_{m})}_{\underline{\mu }_{m}}(\tilde{f} _{1}), \tilde{\Phi } ^{(\underline{\kappa }_{n})}_{\underline{\mu }_{n}}(\tilde{f} _{2}))$$
$$      = (\Phi ^{(\underline{\kappa }_{m})}_{\underline{\mu }_{m}}(f_{1}), \Phi ^{(\underline{\kappa }_{n})}_{\underline{\mu }_{n}}(f_{2})) = {\mathcal W} ^{(\underline{\kappa }_{m+n})}_{\underline{\mu }_{m+n}}(f^{*}_{1} \otimes f_{2})$$
$$      =\langle (2\pi )^{2}\delta (q_{0})\tilde{W} ^{(\underline{\kappa }_{m+n})}_{\underline{\mu }_{m+n}}(q_{1}, \ldots , q_{m+n-1}),$$
$$       \overline{\tilde{g} }_{1}(q_{m} - q_{0}, \ldots , q_{1} - q_{0}) \tilde{g} (q_{m}, \ldots , q_{m+n-1})\rangle $$
$$      = (2\pi )^{2}\langle \tilde{W} ^{(\underline{\kappa }_{m+n})}_{\underline{\mu }_{m+n}}(q_{1}, \ldots , q_{m+n-1}), \overline{\tilde{g} }_{1}(q_{m}, \ldots , q_{1}) \tilde{g} (q_{m}, \ldots , q_{m+n-1})\rangle .$$
This identity implies that the support of $\tilde{W}
^{(\underline{\kappa }_{n})}_{\underline{\mu }_{n}}(q_{1}, \ldots ,
q_{n-1})$ is contained in $\Sigma ^{n-1}$ (see Proposition 4.6 of
[\cite{BN04}]). Moreover, the equality
$$      (\tilde{Z} ^{(\underline{\kappa }_{n})}_{\underline{\mu }_{n}}(\tilde{g} ), \tilde{Z} ^{(\underline{\kappa }_{n})}_{\underline{\mu }_{n}}(\tilde{g} )) $$
$$      = (2\pi )^{2}\langle \tilde{W} ^{(\underline{\kappa }_{2n})}_{\underline{\mu }_{2n}}(q_{1}, \ldots , q_{2n-1}), \overline{\tilde{g} }(q_{n}, \ldots , q_{1}) \tilde{g} (q_{n}, \ldots , q_{2n-1})\rangle $$
shows that the support of $\tilde{Z} ^{(\underline{\kappa }_{n})}_{\underline{\mu }_{n}}(q_{0}, \ldots , q_{n-1})$ is contained in
$\Sigma ^{n}$.  From this support property it follows that $\tilde{Z} ^{(\underline{\kappa }_{n})}_{\underline{\mu }_{n}}(\tilde{g} )$ exists for a much
wider class of test functions $\tilde{g} $ than was originally considered.
For example, the function
$$      \tilde{g} _{\zeta }(q) = (2\pi )^{-2n} e^{ i [\sum _{j=0}^{n-1} q_{j}
 \zeta _{j}]}, \quad  {\rm Im \, }\zeta _{j} \in  V_{+} + \ell _{j} (1, \mbox{\bi 0}) $$
belongs to the class of test functions for sufficiently large $\ell
_{j}$. We investigate the  region of holomorphy of the following
function
$$      \langle \tilde{W} ^{(\underline{\kappa }_{2n})}_{\underline{\mu }_{2n}}(q_{1}, \ldots , q_{2n-1}), \tilde{g} _{\zeta ^{\prime }}^{*}(q_{1}, \ldots , q_{n}) \tilde{g} _{\zeta }(q_{n}, \ldots , q_{2n-1})\rangle $$
$$ =\frac{1}{(2\pi)^{4n}}\langle \tilde{W}^{(\underline{\kappa }_{2n})}
_{\underline{\mu }_{2n}}(q_{1},\ldots, q_{2n-1}), e^{ -i [\sum _{j=1}^{n} q_{n+1-j}
\bar{\zeta } ^{\prime }_{j-1}]} e^{i [\sum _{k=1}^{n} q_{n+k-1}\zeta _{k-1}]}\rangle $$
$$      = W^{(\underline{\kappa }_{2n})}_{\underline{\mu }_{2n}}(-\bar{\zeta } ^{\prime }_{n-1}, \ldots , -\bar{\zeta } ^{\prime }_{0} + \zeta _{0}, \ldots , \zeta _{n-1}).$$
Now, we recall the following proposition.
\begin{prop}[Proposition 4.7 of {[\cite{BN04}]}]

 There exist decreasing functions $R_{ij}(r)$ defined
for $\ell  < r$ such that $W^{(\underline{\kappa }_{2n})}_{\underline{\mu }_{2n}}
(\zeta _{1}, \ldots , \zeta _{2n-1})$ is holomorphic in
\begin{multline*}
  \cup _{i=1}^{2n-1} \{ \zeta \in \mathbb C^{4(2n-1)}; {\rm Im \, }
  \zeta _{i} \in  V_{+} + (\ell ^{\prime }, \mbox{\bi 0}),
  {\rm Im \, }\zeta _{j} \in  V_{+} + (R_{ij}(\ell ^{\prime }),
  \mbox{\bi 0}),\\ \ell  < \ell ^{\prime }, j \neq  i\} .
\end{multline*} \end{prop}
This proposition shows that $Z^{(\underline{\kappa
}_{n})}_{\underline{\mu }_{n}}(\zeta _{0}, \ldots , \zeta _{n-1})$
is holomorphic in the domain  ${\rm Im \, }
  \zeta _{0} \in  V_{+} + (\ell, \mbox{\bi 0})/2$ and ${\rm Im \, }
  \zeta _{k} \in  V_{+} + (\ell _{k}, \mbox{\bi 0})$ for
  sufficiently large $\ell_{k}$ for $k = 1, \ldots , n-1$.
\indent Note that
$$      (\tilde{g} _{\zeta }\circ \chi _{n}^{-1})(p_{1}, \ldots , p_{n}) = (2\pi )^{-2n} \exp i \langle \zeta , \chi _{n}^{-1}p\rangle $$
$$      = (2\pi )^{-2n} \exp i \langle \chi _{n}^{-1T}\zeta , p\rangle  = (2\pi )^{-2n} \exp i \langle z, p\rangle ,$$
where $z = \chi _{n}^{-1T}\zeta $ and $\zeta  = \chi _{n}^{T}z$, that is,
$$      \zeta _{0} = z_{1}, \  \zeta _{j} = z_{j+1} - z_{j} \  (j = 1, \ldots , n-1),$$
$$      z_{1} = \zeta _{0}, \  z_{j} = \sum _{k=0}^{j-1} \zeta _{k} \  (j = 2, \ldots , n).$$
Therefore we get
$$      Z^{(\underline{\kappa }_{n})}_{\underline{\mu }_{n}}(\zeta _{0}, \ldots , \zeta _{n-1}) = \tilde{Z} ^{(\underline{\kappa }_{n})}_{\underline{\mu }_{n}}(\tilde{g} _{\zeta })$$
$$      = \tilde{\Phi } ^{(\underline{\kappa }_{n})}_{\underline{\mu }_{n}}(\tilde{g} _{\zeta }\circ \chi _{n}^{-1}) = \Phi ^{(\underline{\kappa }_{n})}_{\underline{\mu }_{n}}(z_{1}, \ldots , z_{n}),$$
and
$$      \Phi ^{(\underline{\kappa }_{n})}_{\underline{\mu }_{n}}(f)
= \int \Phi ^{(\underline{\kappa }_{n})}_{\underline{\mu
}_{n}}(x_{1} + i\ell _{0}, \ldots , x_{n} + i\sum _{k=1}^{n}\ell
_{k-1})$$
$$      \times  f(x_{1} + i\ell _{0}, \ldots , x_{n} + i\sum _{k=1}^{n}\ell _{k-1}) dx_{1} \cdots dx_{n}, $$
where $\ell _{0} = \ell /2 + \epsilon $ for any $\epsilon > 0$.
Note that the Poincar\'e group acts on $\tilde{g}  _{\zeta }(q)$
as
$$      (a, A): \tilde{g}  _{\zeta }(q) \rightarrow  \tilde{g}  _{\zeta }(\Lambda (A)^{-1}q)e^{iaq_{0}} = (2\pi )^{-2n}e^{i[\sum _{j=0}^{n-1} \Lambda (A)^{-1}q_{j}\zeta _{j}]}e^{iaq_{0}}$$
$$      = (2\pi )^{-2n} e^{i[\sum _{j=0}^{n-1} q_{j}\Lambda (A)\zeta _{j}]}e^{iaq_{0}} = \tilde{g}  _{\Lambda (A)\zeta }(q)e^{iaq_{0}}.$$
Then the formula of covariance
$$      U(a, A)\Phi ^{(\underline{\kappa }_{n})}_{\underline{\mu }_{n}}(f) = \sum _{\nu _{1}, \dots , \nu _{n}} \prod _{j=1}^{n}V^{(\kappa _{j})}_{\mu _{j},\nu _{j}}(A^{-1})\Phi ^{(\kappa _{1} \ldots \kappa _{n})}_{\nu _{1} \ldots \nu _{n}}(f_{(a, A)})$$
implies the following simple formula of covariance in the
 domain of holomorphy of  $\Phi^{(\underline{\kappa
}_{n})}_{\underline{\mu }_{n}}(z_{1}, \ldots , z_{n})$ in
complex space:
$$      U(a, A)\Phi ^{(\underline{\kappa }_{n})}_{\underline{\mu }_{n}}(z_{1}, \ldots , z_{n})$$
$$      = \sum _{\nu _{1}, \dots , \nu _{n}} \prod _{j=1}^{n}V^{(\kappa _{j})}_{\mu _{j},\nu _{j}}(A^{-1})\Phi ^{(\kappa _{1} \ldots \kappa _{n})}_{\nu _{1} \ldots \nu _{n}}(\Lambda (A)z_{1} + a, \ldots , \Lambda (A)z_{n} + a). \eqno{(5.1)}$$

\vskip 12pt \noindent

\section{Multiplication of $\rho(x)$ and $\psi(x)$}
 As stated at the
end of Section 3, $\{ {\mathcal H} , \Phi _{0}, U(a, \Lambda ), \phi
(x), \rho (x), \rho ^{*}(x)\} $ satisfies the axioms of UHFQFT. Let
$\rho ^{(\kappa )}(x) = \rho (x)$ and $\rho ^{(\bar{\kappa })}(x) =
\rho ^{*}(x)$. Then, as we learned in the previous section, the
vector-valued function $\rho ^{(\lambda _{1})}(z_{1}) \cdots \rho
^{(\lambda _{n})}(z_{n})\Phi _{0}$ is holomorphic in
$$ \{ (z_{1}, \ldots , z_{n}) \in  \mathbb C^{4n}; {\rm Im \, }z_{1} \in  V_{+} + (\ell_{0}, \mbox{\bi 0}), {\rm Im \, }(z_{j+1} - z_{j}) \in  V_{+} + (\ell _{j}, \mbox{\bi 0})\} $$
for some $\ell _{j} > \ell  > 0$ $(j = 1, \ldots , n-1)$, where $\rho
^{(\lambda )}(x)$ is one of  $\rho^{(\kappa )}(x)$, $\rho
^{(\bar{\kappa } )}(x)$ and $\phi (x)$.  Let $\psi_{0, \alpha
}^{(\kappa )}(x) = \psi_{0, \alpha }(x)$ and $\psi_{0, \bar{\alpha
} }^{(\bar{\kappa } )}(x) = {\psi }_{0, {\alpha } }^*(x)$ be a
free Dirac fields of mass $M$. Then the system $$\{ {\mathcal K} ,
\Psi _{0}, V(a, \Lambda ), \psi _{0, \alpha }^{(\kappa )}(x), \psi
_{0, \bar{\alpha } }^{(\bar{\kappa } )}(x)\} $$ satisfies the axioms
of tempered field theory (and consequently, that of UHFQFT), and
therefore $\psi ^{(\lambda _{1})}_{0, \beta _{1}}(z_{1}) \cdots \psi
^{(\lambda _{n})}_{0, \beta _{n}}(z_{n})\Psi _{0}$ is holomorphic in
$$      \{ (z_{1}, \ldots , z_{n}) \in  \mathbb C^{4n}; {\rm Im \, }z_{1} \in  V_{+}, {\rm Im \, }(z_{j} - z_{j-1})  \in  V_{+}\} ,$$
where $\lambda  = \kappa $, $\beta  = \alpha $ or $\lambda  =
\bar{\kappa } $, $\beta  = \bar{\alpha } $.  Therefore, $\rho
(z)\Phi $ for $\Phi  = \rho ^{(\lambda _{2})}(f_{2}) \cdots \rho
^{(\lambda _{n})}(f_{n})\Phi _{0}$, $f_{j} \in  {\mathcal T}
(T(\mathbb R^{4}))$ is holomorphic in
$$      \{ z \in  \mathbb C^{4}; {\rm Im \, }z \in  V_{+} + (\ell /2, \mbox{\bi 0})\} $$
and $\psi _{0, \alpha _{1}}(z)\Psi $ for $\Psi  = \psi ^{(\lambda
_{2})}_{0, \beta _{2}}(g_{2}) \cdots \psi ^{(\lambda _{n})}_{0,
\beta _{n}}(g_{n})\Psi _{0}$, $g_{j} \in  {\mathcal S} (\mathbb
R^{4})$ is holomorphic there too.\newline The composite system
\begin{multline*}
\{
{\mathcal H} \otimes {\mathcal K} , \Phi _{0}\otimes \Psi _{0},
U(a,\Lambda )\otimes V(a,\Lambda ), \phi (x)\otimes I_{{\mathcal K} },
\rho (x)\otimes I_{{\mathcal K} },\\ \rho ^{*}(x)\otimes I_{{\mathcal
K} }, I_{{\mathcal H} }\otimes \psi _{0, \alpha }(y), I_{{\mathcal
H} }\otimes \bar{\psi } _{0, \bar{\alpha } }(y)\} \end{multline*} is
the tensor product of two systems and thus satisfies all the axioms of UHFQFT. Although
the tensor product is well-defined, the pointwise product is not necessarily
well-defined for generalized (vector-valued) functions.  In the
category of distributions, the following theorem is well-known:

\begin{thm}[Theorem 8.2.10 of {[\cite{Hoe83}]}]  If $u, v \in {\mathcal D}^{\prime
} (X)$ then the product $uv$ can be defined as the pullback of the
tensor product $u \otimes v$ by the diagonal map $\delta : X
\rightarrow X \times X$ unless $(x, \xi ) \in WF(u)$ and $(x, -\xi )
\in WF(v)$.\end{thm}

In our case, the condition that $\rho (z)\Phi $ and $\psi _{0,
\alpha _{1}}(z)\Psi $ have the common domain of holomorphy,
$$      \{ z \in  \mathbb C^{4}; {\rm Im \, }z \in  V_{+} + (\ell /2, \mbox{\bi 0})\},$$
which corresponds to the condition of the wave front sets $WF(u)$
and $WF(u)$ of distributions, implies that the product $(\psi _{0,
\alpha }\rho )(f)$ is well-defined  by the formula
\begin{multline*}
 (\psi _{0, \alpha }\rho )(f)(\Psi  \otimes  \Phi ) =
 \int _{\Gamma _{N}}f(z)\psi _{0, \alpha }(z)\Psi  \otimes \rho (z)
 \Phi  dz, \\  \Gamma _{N} = \{ z \in  \mathbb C^{4}; z = x + i(N,
 \mbox{\bi 0})\} \end{multline*}
 for suitable $N > 0$.  Thus the field $\psi _{0}(x)$ is a multiplier of
the field $\rho (x)$. Similarly one can show that
$\frac{\partial}{\partial x^{\mu}} \psi_{0,\alpha}$ is a multiplier
for $\rho(x)$ and then we calculate
$$      (\frac{\partial }{\partial x^{\mu }}(\psi _{0, \alpha }\rho ))(f)\Psi  \otimes \Phi $$
$$       = (\psi _{0, \alpha }\rho )(-\frac{\partial }{\partial x^{\mu }}f)\Psi  \otimes \Phi  = \int _{\Gamma _{N}} (-\frac{\partial }{\partial x^{\mu }}f(z)) \psi _{0, \alpha }(z)\Psi  \otimes  \rho (z)\Phi  dz$$
$$       = \int _{\Gamma _{N}} f(z) \{ (\frac{\partial }{\partial x^{\mu }}\psi _{0, \alpha }(z)\Psi ) \otimes  \rho (z)\Phi  + \psi _{0, \alpha }(z)\Psi  \otimes  \frac{\partial }{\partial x^{\mu }}\rho (z)\Phi \}  dz$$
$$       = ((\frac{\partial }{\partial x^{\mu }}\psi _{0, \alpha })\rho )(f)\Psi  \otimes \Phi  + (\psi _{0, \alpha }\frac{\partial }{\partial x^{\mu }}\rho )(f)\Psi  \otimes \Phi .$$
This gives
\begin{multline*}
$$       \frac{\partial }{\partial x^{\mu }}(\psi _{0, \alpha }(x) \rho (x))(\Psi  \otimes \Phi )
     =\\ (\frac{\partial }{\partial x^{\mu }}\psi _{0, \alpha } (x))\rho (x) \Psi  \otimes \Phi
     + \psi _{0, \alpha } (x) \frac{\partial }{\partial x^{\mu }}\rho (x)\Psi
     \otimes \Phi .\end{multline*}
\vskip 12pt \noindent Let $\psi (x) = \psi _{0}(x)\rho (x)$ and
$\bar{\psi } (x) = \bar{\psi } _{0}(x)\rho ^{*}(x)$.  We can
easily see that the fields $\psi (x), \bar{\psi } (x), \phi (x)$
satisfy the axioms of UHFQFT except for the extended causality,
which is proved in Section 2.  In fact, the conditions WI - WV follow from
those of the systems $$\{ {\mathcal H} , \Phi _{0}, U(a, \Lambda ),
\phi (x), \rho (x), \rho ^{*}(x)\} \; \rm{and} \; \{ {\mathcal K} , \Psi
_{0}, V(a, \Lambda ), \psi _{0, \alpha }^{(\kappa )}(x), \psi _{0,
\bar{\alpha } }^{(\bar{\kappa } )}(x)\} $$ (for WV the relation
(5.1) is used).  For WVII, we only have  to restrict the Hilbert
space ${\mathcal H} \otimes {\mathcal K}$ to the subspace
generated by
$$        \phi ^{(\kappa _{1})}_{j_{1}}(f_{1}) \cdots \phi ^{(\kappa _{n})}_{j_{n}}(f_{n}) \Phi _{0} \otimes \Psi _{0}, \  f_{j} \in  {\mathcal T} (T(\mathbb R^{4})) \
(n = 0, 1, \ldots ),$$ where $\phi ^{(\kappa )}_{j}(x)$ is $\psi
_{\alpha }(x) = (I_{{\mathcal H} }\otimes \psi _{0, \alpha
}(x))\cdot (\rho (x)\otimes I_{{\mathcal K} }) = \rho (x)\otimes
\psi _{0, \alpha }(x)$ or $\bar{\psi } _{\bar{\alpha } }(x) =
(\rho ^{*}(x)\otimes I_{{\mathcal K} })\cdot (I_{{\mathcal H} }
\otimes \bar{\psi } _{0, \bar{\alpha } }(x)) = \rho ^{*}(x)\otimes
\bar{\psi } _{0, \bar{\alpha } }(x)$ or $\phi (x) \otimes
I_{{\mathcal K} }$.
\vskip 12pt

At the end of this section we complete the proof of the condition of extended
causality in the form of axiom WVI by showing that this axiom
is equivalent to Condition (R3) for the Wightman functionals which has been verified in Section 2.
\begin{prop}
Assuming the validity of the other axioms, the axiom of extended
causality WVI is equivalent to the following condition
\begin{enumerate}
\item[({\bf R3})]
  For all $n=2, 3, \ldots $ and all $i =
1, \ldots , n-1$ denote
$$  L_{i}^{\ell } = \{ x = (x_{1}, \ldots , x_{n}) \in  \R^{4n}; \vert x_{i} - x_{i+1}\vert _{1} < \ell \} ,$$
$$  W_{i}^{\ell } = \{ x = (z_{1} , \ldots , z_{n}) \in  \C^{4n}; z_{i} - z_{i+1} \in V^{\ell }\} .$$
Then, for any $\ell ^{\prime } > \ell $,
\begin{enumerate}
 \item[(i)] the
functional
$${\mathcal T}  (T(\R^{4n})) \ni  f \rightarrow
{\mathcal W} ^{(\kappa _{1} \ldots \kappa _{n})}_{\mu _{1} \ldots \mu _{n}}(f) \in  \C$$
is extended continuously to ${\mathcal T} (T(L_{i}^{\ell ^{\prime }}))$, and
\item[(ii)]
the functional on ${\mathcal T} (T(\R^{4n}))$
$$  f \rightarrow  {\mathcal W}^{(\kappa_{1} \ldots \kappa_{j} \kappa _{j+1}\ldots \kappa_{n})}_{\mu_{1} \ldots \mu_{j} \mu_{j+1} \ldots \mu _{n}}(f)  +
{\mathcal W}^{(\kappa _{1} \ldots \kappa_{j+1} \kappa_{j+} \ldots \kappa _{n})}_{\mu_{1} \ldots \mu_{j+1} \mu_{j} \ldots \mu_{n}}(f)  \in  \C$$
is extended continuously to ${\mathcal T}(W_{i}^{\ell^{\prime}})$.
\end{enumerate}
\end{enumerate}
\end{prop}
\begin{proof}
Since the spinor/tensor indices do not play a role in this
statement the proof given in [\cite{BN04}] for the scalar case
applies (see Propositions 4.3, 4.4 and Theorem 5.1 of
[\cite{BN04}]).
\end{proof}

\noindent
From the tensor structure of the composite system of $\rho(x)$ and
$\psi_{0}(x)$, the Wightman function of $\psi(x) = \psi_{0}(x)\rho (x)$ is the product of the Wightman functions of $\psi _{0}(x)$ and $\rho (x)$.  Then it follows from Propositon 3.3 that the Wightman functional of $\psi(x)$ is just
(2.1) for which the extended causality (R3) is proven in Section 2.
Thus the axiom WVI is verified.

\section{Conclusion }
After the condition of extended causality had been verified in its functional version (Section 2), this second part of our study of a linearized model of Heisenberg's fundamental equation established first the convergence of the Wick power series
$$\rho(x)=:e^{il^2 \phi(x)^2}:= \sum_{n=0}^{\infty} \frac{(il^2)^n}{n!}
:\phi(x)^{2n}:$$
through Wick power series techniques. It turns out that this power series converges in the sense of tempered ultra-hyperfunctions but not in the sense of (tempered)
Schwartz distributions.

Next through the use of further Wick product techniques it is shown that this field $\rho$ satisfies the differential equation (in the sense of operator-valued tempered ultra-hyperfunctions)
$$\partial_{\mu}\rho(x)=2il^2:\rho(x)\phi(x)\partial_{\mu}\phi(x):$$
where we used the abbreviation $\partial_{\mu}= \frac{\partial}{\partial x^{\mu}}.$

Finally, in order to solve the system (\ref{eq:soluble}) by the ansatz
\beq \label{ansatz}
\psi(x)=\psi_0(x)\rho(x)
\eeq
with $\psi_0$ being a free Dirac field two results have been established, namely
\begin{enumerate}
\item[a)] the concept of a relativistic quantum field with a fundamental length of general type $\kappa$ (i.e., a scalar, tensor or spinor field) generalizing the case of a scalar field presented in [\cite{BN04}] and
  \item[b)] the free Dirac field $\psi_0$ is a multiplier of the field $\rho$.
\end{enumerate}
Then it follows that the field $\psi$ in (\ref{ansatz}) is a relativistic quantum field with a fundamental length of spinor type which satisfies the system (\ref{eq:soluble}). The interpretation and the motivation of our use of the concept
of a quantum field theory with a fundamental length can also be found in the introduction to part I and in [\cite{BN04}].

We find it a very remarkable fact that the length parameter $l$ in the linearized version of Heisenberg's fundamental equation can be interpreted as the fundamental length in the sense of our theory of relativistic quantum field theory with a fundamental length as developed in [\cite{BN04}].

As important physical consequences
we mention that therefore the solution of the linearized version of Heisenberg's fundamental equation falls in the class of quantum field theories for which the PCT and spin-statistic  theorems hold and for which a scattering theory is available.\\[5mm]

\noindent
{\bf Acknowledgements}. This paper is part of the research project Re-
search for axiomatic quantum field theory by using ultra-hyperfunctions,
grant number 16540159, Grants-in-Aid for Scientific Research of JSPS
(Japan Society for the Promotions of Science). The authors gratefully
acknowledge substantial financial support by the JSPS.
Major parts of the work for this article were done during a research
visit of E. B. to the University of Tokushima, funded by this grant.
With great pleasure, E. B. expresses his gratitude to the Department
of Mathematics, in particular his host S. N., and the JSPS.

\bibliographystyle{plain}
\bibliography{hfqft20071224}

\end{document}